\documentclass{optica-article}
\journal{opticajournal} %
\articletype{Research Article}
\usepackage{lineno}

\usepackage{bm}%
\usepackage{physics}
\usepackage{mathtools}

\usepackage{cleveref}
\crefname{equation}{Eq.}{Eqs.} %
\crefrangelabelformat{equation}{(#3#1#4--#5#2#6)}

\usepackage{soul}
\usepackage{cancel}%

\usepackage[normalem]{ulem}   %

\soulregister\bm        7    %
\soulregister\mathrm    7
\soulregister\cancel    7
\soulregister\textcolor 7    %
\soulregister\sout 7    %
\soulregister\cite 7
\soulregister\ref 7
\soulregister\color 7 

\newcommand*\linenomathpatch[1]{%
  \cspreto{#1}{\linenomath}%
  \cspreto{#1*}{\linenomath}%
  \csappto{end#1}{\endlinenomath}%
  \csappto{end#1*}{\endlinenomath}%
}

\linenomathpatch{equation}
\linenomathpatch{gather}
\linenomathpatch{multline}
\linenomathpatch{align}
\linenomathpatch{alignat}
\linenomathpatch{flalign}

\begin{document}
\title{Wave--particle duality of unpolarized photons}

\author{Naofumi Abe\authormark{1,*} and Keiichi Edamatsu\authormark{1,2}}

\address{\authormark{1}Research Institute of Electrical Communication, Tohoku University, Sendai, 980-8577, Japan

\authormark{2}Present address: Department of Physics, Tohoku University, Sendai,
 980-8578, Japan}

\email{\authormark{*}abe.naofumi.a56@kyoto-u.jp} %

\begin{abstract*}
Photons in a two-path interferometer best embody wave--particle duality (WPD), which is a core concept of quantum theory. 
So far, the WPD relation is commonly written as $V^2+D^2 \leq 1$, where $V$ is the interference fringe visibility and $D$ is path distinguishability, i.e., the distinguishability of which path a photon passed.
This inequality is saturated only when the which-way marker (WWM), which embodies which-path information (WPI) via an internal degree of freedom of photons, such as polarization, is in a pure state.
For mixed-state WWM, conventionally defined distinguishability underestimates the amount of WPI and thus does not saturate the WPD relation. 
Here, we introduce a generalized measure of distinguishability $D$ that properly quantifies the WPI and saturates the WPD relation for all pure- and mixed-state WWM within a purification-based framework. 
To this end, mixed-state WWM is treated as a result of entanglement formation between the WWM and an external degree of freedom, e.g., environment, and $D$ is defined so that it incorporates the total WPI shared between the WWM and the environment.
We show that $D$ thus defined is experimentally quantifiable, independently of $V$, without access to the environment.
We experimentally evaluate $V$ and $D$ using true single photons generated in the completely mixed (unpolarized) state, and thus verify the saturated WPD relation.
\end{abstract*}

\section{Introduction} 
Quantum interference and wave--particle duality (WPD) are crucial properties of quantum physics. 
Various quantum objects
exhibit complementary behaviors: wave and particle \cite{jacques2005, xiao2019observing, bach2013, zeilinger1988, carnal1991, nairz2003}. 
\textcolor{black}{
The wave nature is characterized by {\it interference} and thus the
}%
degree of wave 
\textcolor{black}{
nature
}%
is quantified by the interference fringe visibility $V$. 
\textcolor{black}{
On the other hand, the particle nature is characterized by 
the {\it locality} of a single object.
}%
In the case of two-path interferometers, 
which-path information (WPI)
\textcolor{black}{
indicates the knowledge about the locality (path) of the object.
}%
An inner degree of freedom of 
\textcolor{black}{
the quantum object,
}%
such as polarization of a photon, is used to add 
the WPI
and the inner degree of freedom is 
called a which-way marker (WWM). 
\textcolor{black}{
The degree of WPI and thus the degree of particle nature 
}%
is quantified by the path distinguishability \textcolor{black}{$D$}, 
\textcolor{black}{
which
}%
is conventionally 
\textcolor{black}{quantified by} half of the trace distance between 
the WWM states in each path mode \cite{englert1996, englert1999}. 
\textcolor{black}{
It is known that $V$ and 
\textcolor{black}{$D$} 
follow
}%
the complementary relation, 
i.e., the WPD relation \textcolor{black}{or}
the WPD inequality \cite{englert1996, englert1999, bosyk2013, zela2018, jara2022experimental, sanchez2019}, %
\begin{equation}
V^2+
\textcolor{black}{D}^2
 \leq 1. 
\label{eq:wpd}
\end{equation}
This inequality has been experimentally verified using photons \cite{schwindt1999, liu2012, jacques2007, jacques2008delayed, kaiser2012, ma2016, steuernagel2007, jacques2008illustration} and atoms \cite{durr1998}. 
Equation~(\ref{eq:wpd}) 
\textcolor{black}{
is saturated,
}%
i.e., 
\textcolor{black}{
\begin{equation}
V^2+
D^2 
= 1 
\label{eq:WPD-pure}
\end{equation}
}%
holds when the WWM is in pure states \cite{englert1996, englert1999, schwindt1999, zela2018, sanchez2019}. 

However, for WWM in mixed states, 
\textcolor{black}{
$D_\mathrm{c}$
(hereafter we refer to conventionally defined $D$ for mixed-state WWM as $D_\mathrm{c}$)
does not saturate the WPD inequality Eq.~(\ref{eq:wpd}),
implying that $D_\mathrm{c}$ 
somehow underestimates the 
\textcolor{black}{total} WPI 
\textcolor{black}{for} mixed-state WWM.
}%
For instance, 
in completely unpolarized single photons, each photon 
still 
\textcolor{black}{
perfectly interferes
}%
with itself, 
even though 
\textcolor{black}{
there is
}%
no coherence with respect to polarization \cite{ericsson2005, sanchez2019}. 
\textcolor{black}{
Even in this case, it is known that
}%
polarization rotation 
\textcolor{black}{
in
}%
either path 
\textcolor{black}{
degrades the interference visibility
}%
$V$ \cite{sanchez2019},
\textcolor{black}{
indicating that 
}%
WPI is added in the polarization
\textcolor{black}{
degrees of freedom even in the completely unpolarized state.
}%
In this situation, 
$D_\mathrm{c}$ always gives zero 
\cite{schwindt1999}, 
\textcolor{black}{
because
}%
the density matrix of the unpolarized state 
is invariant to any polarization rotation. 
Thus, as exemplified above for the completely unpolarized state, 
for WWM in mixed states the conventional path distinguishability $D_\mathrm{c}$ is not a good measure of the distinguishability that, together with the visibility $V$, should %
\textcolor{black}{%
saturate 
}%
\textcolor{black}{
the WPD inequality
}%
Eq.~(\ref{eq:wpd}).

Meanwhile,
another duality relation  (See Eqs.~(\ref{eq:general.pct})--(\ref{eq:general.pct-c0}))
\begin{equation}
V^2+C^2=1 
\label{eq:pct-c}
\end{equation}
is shown in the case of WWM scenario,
where $C$ is the concurrence, a measure of entanglement between the path qubit and the WWM.
Combining Eqs.~(\ref{eq:WPD-pure}) and (\ref{eq:pct-c}), 
we see that $D$ and $C$ are equivalent
in the case where the total system is in a pure state \cite{Qureshi2021Predictability}. 
Thus, in principle, $C$ can be another measure of distinguishability.
However, in practice, it is difficult to evaluate $C$ independently of $V$. 
Nevertheless, it is worth observing the close relation between the distinguishability and the entanglement in the WWM scenario.

In this paper, we introduce a generalized measure of path distinguishability $D$, so that it always satisfies the WPD equality, i.e., Eq.~(\ref{eq:WPD-pure}), even in mixed WWM states as in pure states \textcolor{black}{within a purification-based framework}. 
To this end, we attribute mixed WWM states to entanglement formation with other degrees of freedom, e.g., environment, 
and show that WPI
in this case is shared between the WWM and the environment.
Thus, $D$ is defined so that it incorporates WPI held not only by the WWM itself but also by the environment.    
We also show how we can experimentally quantify $D$ even in usual situations where we cannot access the environment that is entangled with the WWM.
Then, we experimentally verify 
\textcolor{black}{
$V$ and $D$
}%
for unpolarized 
\textcolor{black}{true} 
single photons 
\textcolor{black}{emitted from a nitrogen-vacancy (NV) center in diamond \cite{abe2017}}
and demonstrate  
\textcolor{black}{
that they almost 
}%
\textcolor{black}{
agree with the 
WPD equality 
\textcolor{black}{
Eq.~(\ref{eq:WPD-pure}),
}%
}%
\textcolor{black}{
as predicted. 
}%
Furthermore, 
we show that 
\textcolor{black}{
WPI-induced 
}%
destruction and 
\textcolor{black}{
quantum-erasure (erasure of the WPI)-induced
}%
revival of 
\textcolor{black}{
the
}%
interference are also possible for unpolarized 
single 
photons,
as in the case of 
photons
\textcolor{black}{
in pure polarization states.
}%

\section{Theory}\label{Sec.Theory}%
\subsection{General consideration}\label{sec:general}%
We consider a general two-path interferometer system (Q) interacted with another system (R).
The total system (QR) is assumed to be in a pure state $\ket{\Psi}_\mathrm{QR}$ described by the entangled form 
\begin{equation}
\ket{\Psi}_\mathrm{QR}
= c_0 \ket{0}_\mathrm{Q}\otimes\ket{\psi_0}_\mathrm{R}  
+ c_1 \ket{1}_\mathrm{Q}\otimes\ket{\psi_1}_\mathrm{R} ,  
\label{eq:general.statevec}
\end{equation}
where $\ket{0}_\mathrm{Q}$ and $\ket{1}_\mathrm{Q}$ are the orthonormal bases of the path qubit of the interferometer,
$\ket{\psi_0}_\mathrm{R}$ and $\ket{\psi_1}_\mathrm{R}$ are the normalized state vectors of the system R coupled to $\ket{0}_\mathrm{Q}$ and $\ket{1}_\mathrm{Q}$, respectively,
and $|c_0|^2+|c_1|^2=1$.
Note that the system R is not restricted to a qubit system; it may be a multidimensional or a multipartite system.
In Eq.~(\ref{eq:general.statevec}), the two-dimensional subspace in R 
spanned by $\ket{\psi_0}_\mathrm{R}$ and $\ket{\psi_1}_\mathrm{R}$
is used. 
The reduced density operator $\rho_\mathrm{Q}$ of the system Q is given by
\begin{alignat}{1}
\rho_\mathrm{Q}
&=\Tr_\mathrm{R} \ket{\Psi}\bra{\Psi} \nonumber\\
&= |c_0|^2 \ket{0}\bra{0}  
+ |c_1|^2 \ket{1}\bra{1} + c_0c_1^* \braket{\psi_1}{\psi_0}\ket{0}\bra{1}
+ c_1c_0^* \braket{\psi_0}{\psi_1}\ket{1}\bra{0}
 ,  
\label{eq:general.rdoq}
\end{alignat}
where the subscripts (Q, R, QR) of the state vectors are omitted for simplicity. 
From $\rho_\mathrm{Q}$, 
one can obtain the interference visibility $V$, 
the predictability $D_\mathrm{P}$,  
and the degree of (path-mode) polarization 
$P$ 
as \cite{eberly2017, qian2018, zela2018, Kanseri2018Experimental, Qian2020Turning} 
\begin{alignat}{2}
&V
&&= \left|2c_0 c_1\braket{\psi_0}{\psi_1}\right|, \label{eq:general.vis}\\
&D_\mathrm{P}
&&= \left||c_0|^2-|c_1|^2\right|, \label{eq:general.dp}\\
&P
&&=\sqrt{2\gamma_\mathrm{Q}-1}
=\sqrt{1-|2c_0c_1|^2+|2c_0c_1\braket{\psi_0}{\psi_1}|^2},
\label{eq:general.pq}
\end{alignat}
respectively,
where 
$\gamma_\mathrm{Q}=\mathrm{Tr}\rho_\mathrm{Q}^2$ 
is the purity of the path qubit. 
Equations~(\ref{eq:general.vis})--(\ref{eq:general.pq}) lead to 
the polarization coherence theorem (PCT) \cite{eberly2017, qian2018, zela2018, Kanseri2018Experimental, Qian2020Turning}: 
\begin{equation}
V^2+D_\mathrm{P}^2=P^2
.
\label{eq:general.pct}
\end{equation}
Here,
$D_\mathrm{P}$ (also referred to as `distinguishability' in the original literature of PCT\cite{eberly2017, qian2018, zela2018, Kanseri2018Experimental, Qian2020Turning}) 
quantifies in part the which-way information through the intensity difference between the two paths.
$P$ in Eq.~(\ref{eq:general.pq}) is a function of the purity $\gamma_\mathrm{Q}$, which in turn connects to amount of entanglement existing in the total system Eq.~(\ref{eq:general.statevec})
so that
\begin{equation}
C
=\sqrt{2\left(1-\gamma_\mathrm{Q}\right)}, 
\label{eq:general.cq}
\end{equation}
where 
$C$
is the concurrence, an entanglement monotone
\textcolor{black}{
of the state $\ket{\Psi}_\mathrm{QR}$.
}%
Although 
$C$
is originally defined in two-qubit systems  \cite{Hill1997Entanglement, Wootters1998Entanglement},
Eq.~(\ref{eq:general.cq}) is given also by
the generalized version of concurrence known as I-concurrence \cite{Rungta2001Universal, Torre2022Relationship, Marrou2023Wave, Qian2020Quantification}, which is applicable to multidimensional or multipartite systems.
Using Eqs.~(\ref{eq:general.pq}) and (\ref{eq:general.cq}), 
the complementary relation 
\begin{equation}
C^2+P^2=1
\label{eq:general.dop-c}
\end{equation}
is shown \cite{Qian2016Coherence}. 
By combining Eqs.~(\ref{eq:general.pct}) and (\ref{eq:general.dop-c}), 
the visibility--predictability--concurrence triality relation, 
\begin{equation}
V^2+D_\mathrm{P}^2+C^2=1{,} 
\label{eq:general.pct-c}
\end{equation}
was obtained \cite{qian2018, Qian2020Turning}. 
In Eq.~(\ref{eq:general.pct-c}),
we see that
$D_\mathrm{P}^2$ 
and 
$C^2$ 
together hold the which-way information 
so that the sum of them is complementary to $V^2$.
Note that the WWM scenario we deal with in this paper takes $|c_0|^2=|c_1|^2=1/2$, thus $D_\mathrm{P}=0$.
In this case, 
the triality relation Eq.~(\ref{eq:general.pct-c}) reduces to the duality relation
\begin{equation}
V^2+C^2=1{.} 
\label{eq:general.pct-c0}
\end{equation}
In the WWM scenario, the distinguishability $D$ is quantified by half of the trace distance between 
$\ketbra{\psi_0}{\psi_0}$ and $\ketbra{\psi_1}{\psi_1}$ 
\cite{englert1996, englert1999}:  
\begin{alignat}{1}
D 
&\coloneqq{}%
\frac12\Tr \bigl| \ketbra{\psi_0}{\psi_0} - \ketbra{\psi_1}{\psi_1} \bigr| \label{eq:general.d} \\
&=
\sqrt{1 - \left| \braket{\psi_0}{\psi_1} \right|^2} ,
\label{eq:general.d1} 
\end{alignat}
where $\Tr|A|\coloneqq \Tr\sqrt{A^\dagger A}$ is the trace norm of matrix $A$.
Combining Eq.~(\ref{eq:general.d1}) with Eq.~(\ref{eq:general.vis})
taking $|c_0|=|c_1|=1/\sqrt2$, we get
\begin{alignat}{1}
V^2+D^2 &=1 ,
\label{eq:general.wpd} 
\end{alignat}
i.e\textcolor{black}{.}, the WPD equality
\textcolor{black}{
seen already in Eq.~(\ref{eq:WPD-pure}).
}%
From Eqs.~(\ref{eq:general.pct-c0}) and (\ref{eq:general.wpd}), 
we see that 
$C$ 
and $D$ are equivalent in this case;
each of them equally holds the which-way information.
However, $D$ has an advantage when we need to evaluate the which-way information experimentally. 
In fact, as we demonstrate later, $D$ can be evaluated with a realistic method 
even when we have limited access to a part of the system R.
In the following, we examine a couple of practical cases in the WWM scenario.

In the first case, we deal with the WWM in pure qubit states:
\begin{alignat}{1}
\ket{\psi_0}_\mathrm{R} &= \ket{\phi_0}_\mathrm{W}, \\
\ket{\psi_1}_\mathrm{R} &= \ket{\phi_1}_\mathrm{W} = U_\mathrm{}\ket{\phi_0}_\mathrm{W}, 
\label{eq:general.purewwm} 
\end{alignat}
where the subscript W stands for WWM, and $U_\mathrm{}$ is the unitary operator that 
\textcolor{black}{transforms} $\ket{\phi_0}_\mathrm{W}$ to $\ket{\phi_1}_\mathrm{W}$ depending on the interferometer paths.
Using Eqs.~(\ref{eq:general.vis}), (\ref{eq:general.d}), and (\ref{eq:general.d1}), $V$ and $D$ are obtained as 
\begin{alignat}{2}
&V
&&= \left|\bra{\phi_0}U_{}\ket{\phi_0}\right|,\\%
&D
&&=
\frac12\Tr \bigl| \ketbra{\phi_0}{\phi_0} - \ketbra{\phi_1}{\phi_1} \bigr| 
= \sqrt{1-\left|\braket{\phi_0}{\phi_1}\right|^2} 
= \sqrt{1-\left|\bra{\phi_0}U_{}\ket{\phi_0}\right|^2}, 
\label{eq:general.vdwpure}
\end{alignat}
where the subscript W is omitted for simplicity. 
It is apparent that the WPD equality 
Eq.~(\ref{eq:general.wpd}) still holds in this case. %
To evaluate the amount of $D$ experimentally,
we can use the projective measurement $\pi_\mathrm{D}$:
\begin{alignat}{1}
\pi_{\mathrm{D}}
&=
\frac12 \bigl( \ketbra{\phi_+}{\phi_+} - \ketbra{\phi_-}{\phi_-} \bigr),
\label{eq:general.opd} 
\end{alignat}
where $\ket{\phi_+}$ ($\ket{\phi_-}$) is the eigenvector
corresponding to the eigenvalue $D$ ($-D$) of 
\textcolor{black}{
$\pi_0-\pi_1$, where $\pi_0\coloneqq \ketbra{\phi_0}{\phi_0}$ and $\pi_1\coloneqq \ketbra{\phi_1}{\phi_1}$.
}%
With $\pi_\mathrm{D}$, $D$ is obtained as
\begin{alignat}{1}
D
&=\frac12\Tr \bigl| \pi_0 - \pi_1 \bigr| 
\label{eq:general.opd2} \\
&=\Tr \pi_{\mathrm{D}} \bigl( \pi_0 - \pi_1 \bigr) \\
&= \frac12 \bigl( \bra{\phi_+}\pi_{0}\ket{\phi_+} -\bra{\phi_-}\pi_{0}\ket{\phi_-} 
- \bra{\phi_+}\pi_{1}\ket{\phi_+} + \bra{\phi_-}\pi_{1}\ket{\phi_-} \bigr), \\
&= \bra{\phi_+}\pi_{0}\ket{\phi_+} + \bra{\phi_-}\pi_{1}\ket{\phi_-} -1
.
\label{eq:general.opdm} 
\end{alignat}
To obtain Eq.~(\ref{eq:general.opdm}), 
the relation $\bra{\phi_+}\rho\ket{\phi_+} + \bra{\phi_-}\rho\ket{\phi_-} =1$
is used, where $\rho$ is any density operator in W.
Since $\left\{ \ket{\phi_+}, \ket{\phi_-}\right\}$ is the orthonormal basis set that gives the maximum outcome ($D$) in Eq.~(\ref{eq:general.opdm}), 
it can be shown that 
Eq.~(\ref{eq:general.opdm}) is equivalent to 
\begin{equation}
D=
\max_{\bm{w}} \left[2 L(\bm{w},\pi_0,\pi_1) -1 \right] ,
\label{eq:general.d_kmax}
\end{equation}
where 
$\bm{w}=\left\{ \ket{w_+}, \ket{w_-}\right\}$ is an orthonormal basis set in W, and
\begin{equation}
L (\bm{w},\pi_0,\pi_1)
\coloneqq 
\frac12 \bigl(
\max \left\{\bra{w_+}\pi_0 \ket{w_+}, \bra{w_+}\pi_1  \ket{w_+} \right\}  
+ \max \left\{\bra{w_-}\pi_0  \ket{w_-}, \bra{w_-}\pi_1 \ket{w_-} \right\} 
\bigr)
\label{eq:general.likelihood_c} 
\end{equation}
is the 
likelihood 
of which-way guess
\cite{englert1996, englert1999, schwindt1999} 
obtained from a certain set of projective measurements  
into $\bm{w}$ %
on the WWM states $\pi_0$ and $\pi_1$.
\textcolor{black}{%
Here, 
\textcolor{black}{
$L$ 
}%
takes 
\textcolor{black}{
$1/2\leq L \leq1$ 
depending 
}%
on $\bm{w}$. 
\textcolor{black}{
$L=1/2$ 
}%
indicates a random guess, and 
\textcolor{black}{
$L=1$ 
}%
indicates that the path is known with certainty.
}%
Thus, 
using the optimized measurement basis set of the WWM in 
Eq.~(\ref{eq:general.opdm}) or (\ref{eq:general.d_kmax}),  
we can experimentally obtain the amount of $D$.
\textcolor{black}{
Note that $1-\max_{w} \{L\}=(1-D)/2$ is also known as the Helstrom bound, 
the minimum-error probability of discrimination between two pure states. 
}%

In the second case, we deal with the WWM in mixed qubit states.
In order to express the mixed-state WWM within the general formalism Eq.~(\ref{eq:general.statevec}),
we consider the entangled pure states for the system R:
\begin{alignat}{1}
\ket{\psi_0}_\mathrm{R} 
&= c_\alpha \ket{\phi_{\alpha0}}_\mathrm{W}\otimes\ket{\alpha}_\mathrm{E}
 + c_\beta \ket{\phi_{\beta0}}_\mathrm{W}\otimes\ket{\beta}_\mathrm{E} , 
\label{eq:general.mixwwm} 
\\
\ket{\psi_1}_\mathrm{R} 
= (U_\mathrm{}\otimes I_\mathrm{E}) \ket{\psi_0}_\mathrm{R}
&= c_\alpha \ket{\phi_{\alpha1}}_\mathrm{W}\otimes\ket{\alpha}_\mathrm{E}
 + c_\beta \ket{\phi_{\beta1}}_\mathrm{W}\otimes\ket{\beta}_\mathrm{E}, 
\label{eq:general.mixwwm1} 
\end{alignat}
where the subscript W stands for the WWM, 
E stands for 
the system (e.g., environment) 
entangled with W, 
$\bigl\{\ket{\alpha}_\mathrm{E}, \ket{\beta}_\mathrm{E}\bigr\}$  
is an orthonormal basis set in E,
$\ket{\phi_{\alpha0}}_\mathrm{W}$ and $\ket{\phi_{\beta0}}_\mathrm{W}$
represent pure WWM states (not necessarily orthogonal to each other) in W,  
$U_\mathrm{}$ is the unitary operator that transforms the WWM state,
$I_\mathrm{E}$ is the identity operator on E,
$\ket{\phi_{\alpha1,\beta1}}_\mathrm{W}=U\ket{\phi_{\alpha0,\beta0}}_\mathrm{W}$,
and $|c_\alpha|^2+|c_\beta|^2=1$.
Note that E may be a multidimensional or a multipartite system; 
$\ket{\alpha}_\mathrm{E}$ and $\ket{\beta}_\mathrm{E}$ 
are two orthogonal vectors that span the two-dimensional subspace in E.
The corresponding reduced density operators $\rho_\mathrm{0}$ and $\rho_\mathrm{1}$ of the WWM states are:
\begin{alignat}{1}
\rho_\mathrm{0} 
&= \Tr_\mathrm{E} \ketbra{\psi_0}
=|c_\alpha|^2\ketbra{\phi_{\alpha0}} + |c_\beta|^2\ketbra{\phi_{\beta0}}, \label{eq:general.mixwwm.rho0} 
\\
\rho_\mathrm{1} 
&= \Tr_\mathrm{E} \ketbra{\psi_1}
=|c_\alpha|^2\ketbra{\phi_{\alpha1}} + |c_\beta|^2\ketbra{\phi_{\beta1}}, \label{eq:general.mixwwm.rho1} 
\end{alignat}
where the subscripts R and W are omitted for simplicity.
Note that Eq.~(\ref{eq:general.mixwwm}) is not a unique expression of the entangled state 
$\ket{\psi_0}_\mathrm{R}$;
we can use the equivalent expression 
\begin{alignat}{1}
\ket{\psi_0}_\mathrm{R} 
&= c_{\alpha'} \ket{\phi_{\alpha'0}}_\mathrm{W}\otimes\ket{\alpha'}_\mathrm{E}
 + c_{\beta'} \ket{\phi_{\beta'0}}_\mathrm{W}\otimes\ket{\beta'}_\mathrm{E}
\label{eq:general.mixwwm2} 
\end{alignat}
using the unitary transformation from the orthonormal basis set 
$\left\{\ket{\alpha}_\mathrm{E},\ket{\beta}_\mathrm{E}\right\}$
to
$\left\{\ket{\alpha'}_\mathrm{E},\ket{\beta'}_\mathrm{E}\right\}$.
Accordingly, for a given $\rho_0$,
one can decompose $\rho_0$ into the mixture of a set of pure states such like 
$\left\{\ket{\phi_{\alpha0}},\ket{\phi_{\beta0}}\right\}$ or $\left\{\ket{\phi_{\alpha'0}},\ket{\phi_{\beta'0}}\right\}$.
Of particular interest is the choice of 
$\left\{\ket{\alpha}_\mathrm{E},\ket{\beta}_\mathrm{E}\right\}$
(and thus $\left\{\ket{\phi_{\alpha0}},\ket{\phi_{\beta0}}\right\}$) 
under which 
$\braket{\phi_{\alpha0}}{\phi_{\alpha1}}=\braket{\phi_{\beta0}}{\phi_{\beta1}}$ holds, 
as we will see later.
Since $\ket{\psi_0}_\mathrm{R}$ and $\ket{\psi_1}_\mathrm{R}$ are pure states,
$V$, 
$C$, and $D$ are readily obtained from 
Eqs.~(\ref{eq:general.vis}), (\ref{eq:general.pct-c0}), %
and (\ref{eq:general.d1}) as follows: 
\begin{alignat}{2}
&V
&&= \left|\braket{\psi_0}{\psi_1}\right| \\
&&&=\left| |c_\alpha|^2 \braket{\phi_{\alpha0}}{\phi_{\alpha1}}
+ |c_\beta|^2 \braket{\phi_{\beta0}}{\phi_{\beta1}} \right|,\\
&C
&&=\sqrt{1-V^2}, \\ 
&D 
&&=\sqrt{1 -V^2}, 
\label{eq:general.mix.vcd} 
\end{alignat}
satisfying the WPD equality Eq.~(\ref{eq:general.wpd}).
To evaluate the amount of $D$ experimentally,
we can use the projective measurement $\Pi_\mathrm{D}$:
\begin{alignat}{1}
\Pi_{\mathrm{D}}
&=
\frac12 \bigl( \ketbra{\psi_+}{\psi_+} - \ketbra{\psi_-}{\psi_-} \bigr),
\label{eq:general.mix.opd} 
\end{alignat}
where $\ket{\psi_+}$ ($\ket{\psi_-}$) is the eigenvector 
corresponding to the eigenvalue $D$ ($-D$)  
of 
\textcolor{black}{
$\Delta\coloneqq\ketbra{\psi_0}{\psi_0} - \ketbra{\psi_1}{\psi_1}$.
Note that the eigenvalues 
of 
$\Delta$
other than $\pm D$ are all zero.
}%
With $\Pi_\mathrm{D}$, $D$ is obtained as
\begin{alignat}{1}
D
&=\Tr \bigl(\Pi_{\mathrm{D}} \Delta \bigr) \\
&= \bigl( \langle{\psi_+}\ketbra{\psi_0}{\psi_0}{\psi_+}\rangle 
+ \langle{\psi_-}\ketbra{\psi_1}{\psi_1}{\psi_-}\rangle -1 \bigr).
\label{eq:general.mix.opdm} 
\end{alignat}
Thus, in principle, by applying $\Pi_\mathrm{D}$ to $\ketbra{\psi_0}{\psi_0}$ and $\ketbra{\psi_1}{\psi_1}$,
we can experimentally obtain the amount of $D$.
However, in general, $\ket{\psi_+}$ and $\ket{\psi_-}$ have entangled forms such as
\begin{alignat}{1}
\ket{\psi_\pm} 
&= d_{\alpha} \ket{\varphi_{\alpha\pm}}_\mathrm{W}\otimes\ket{\alpha}_\mathrm{E}
 + d_{\beta} \ket{\varphi_{\beta\pm}}_\mathrm{W}\otimes\ket{\beta}_\mathrm{E},
\label{eq:general.mix.ev} 
\end{alignat}
where 
$|d_\alpha|^2+|d_\beta|^2=1$
and 
$|d_\alpha|^2\braket{\varphi_{\alpha+}}{\varphi_{\alpha-}} + |d_\beta|^2\braket{\varphi_{\beta+}}{\varphi_{\beta-}}=0$.
Note that $\ket{\varphi_{\alpha+}}$ and $\ket{\varphi_{\alpha-}}$ 
($\ket{\varphi_{\beta+}}$ and $\ket{\varphi_{\beta-}}$)
are not necessarily orthogonal to each other.
It is practically difficult (even impossible in many cases) 
to implement such 
\textcolor{black}{
collective 
}%
operators projective to the entangled states.
Instead, one might consider to use 
the combination of 
\textcolor{black}{
joint operators
}%
\begin{alignat}{2}
&M
&&\coloneqq{}
 \frac12 (M_+ - M_-),
\label{eq:general.mix.lop} 
\\
&M_\pm 
&&\coloneqq{}
\ketbra{\phi_{\alpha\pm}}_\mathrm{W}\otimes\ketbra{\alpha}_\mathrm{E}
+
\ketbra{\phi_{\beta\pm}}_\mathrm{W}\otimes\ketbra{\beta}_\mathrm{E},
\label{eq:general.mix.lop2} 
\end{alignat}
where
$\ket{\phi_{\alpha\pm}}$
($\ket{\phi_{\beta\pm}}$)
are the eigenvectors of 
$\ketbra{\phi_{\alpha0}}-\ketbra{\phi_{\alpha1}}$
($\ketbra{\phi_{\beta0}}-\ketbra{\phi_{\beta1}}$)
having eigenvalues $\pm D_\alpha$ ($\pm D_\beta$).
However, 
$\Tr (M\Delta)\le D$ 
in general,
and the equality holds only if 
$\ket{\phi_{\alpha\pm}}=\ket{\varphi_{\alpha\pm}}$
and
$\ket{\phi_{\beta\pm}}=\ket{\varphi_{\beta\pm}}$.
We find that 
both conditions are met
when
$\braket{\phi_{\alpha0}}{\phi_{\alpha1}}=\braket{\phi_{\beta0}}{\phi_{\beta1}}$.
In other words, 
choosing the bases 
$\left\{\ket{\alpha}_\mathrm{E}, \ket{\beta}_\mathrm{E}\right\}$
under which 
$\braket{\phi_{\alpha0}}{\phi_{\alpha1}}=\braket{\phi_{\beta0}}{\phi_{\beta1}}$
holds, 
we get
\begin{alignat}{1}
D
&=\Tr \bigl( M \Delta \bigr) \\
&= |c_{\alpha}|^2 \bigl( 
 \bra{\phi_{\alpha+}}{\pi_{\alpha0}}\ket{\phi_{\alpha+}} 
 + \bra{\phi_{\alpha-}}{\pi_{\alpha1}}\ket{\phi_{\alpha-}}
 -1 \bigr) \nonumber \\
&\quad + |c_{\beta}|^2\bigl( 
 \bra{\phi_{\beta+}}{\pi_{\beta0}} \ket{\phi_{\beta+}} 
 +\bra{\phi_{\beta-}}{\pi_{\beta1}} \ket{\phi_{\beta-}}
 -1 \bigr), 
\label{eq:general.mix.opdm2} 
\end{alignat}
where
$\pi_{\alpha0} = \ketbra{\phi_{\alpha0}}$,
$\pi_{\alpha1} = \ketbra{\phi_{\alpha1}}$,
etc.
Of course, the value of $D$ thus obtained satisfies the WPD equality in Eq.~(\ref{eq:general.wpd}).
Note, however, any other choice of combination of local projective operators in E and W gives the outcome smaller than $D$.
Thus, 
Eq.~(\ref{eq:general.mix.opdm2}) is equivalent to 
\begin{equation}
D=
\max_{\bm{\epsilon}}\left\{
|c_{\alpha}|^2\max_{\bm{w}_\alpha} \left[2 L(\bm{w}_\alpha,\pi_{\alpha0},\pi_{\alpha1}) -1 \right]
+|c_{\beta}|^2\max_{\bm{w}_\beta} \left[2 L(\bm{w}_\beta,\pi_{\beta0},\pi_{\beta1}) -1 \right]
\right\} ,
\label{eq:general.mix.d_kmax}
\end{equation}
where 
$\bm{\epsilon}=\left\{ \ket{\alpha}_\mathrm{E}, \ket{\beta}_\mathrm{E} \right\}$ 
\textcolor{black}{
($\bm{w}_{\alpha(\beta)}=\left\{ \ket{w_{\alpha(\beta)+}}_\mathrm{W}, \ket{w_{\alpha(\beta)-}}_\mathrm{W} \right\}$) 
}%
is the orthonormal basis set in E (W), and
$L$ is the likelihood defined in Eq.~(\ref{eq:general.likelihood_c}).
Thus,  
using Eq.~(\ref{eq:general.mix.d_kmax}) and the optimized measurement basis sets in both E and W, 
we can experimentally obtain the amount of $D$.
Unlike the case of pure-state WWM in Eq.~(\ref{eq:general.d_kmax}), 
where we need to optimize the measurement set only for W,
in Eq.~(\ref{eq:general.mix.d_kmax}), we must seek the optimized measurement sets for both E and W 
because in this case the which-way information is shared by E and W through the entanglement between them.
Nevertheless, 
in order to evaluate $D$ in Eq.~(\ref{eq:general.mix.opdm2}) or (\ref{eq:general.mix.d_kmax}) in practice,
one \textcolor{black}{does} not have to access the system E, which may be inaccessible in usual cases.
Instead, for a given $\rho_0$ and $\rho_1$, 
one can consider the decomposition into 
the mixture of a set of pure states as given in 
\textcolor{black}{Eqs.~(\ref{eq:general.mixwwm.rho0}) and (\ref{eq:general.mixwwm.rho1})} 
so that 
$\braket{\phi_{\alpha0}}{\phi_{\alpha1}}=\braket{\phi_{\beta0}}{\phi_{\beta1}}$, i.e., 
$\bra{\phi_{\alpha0}}U\ket{\phi_{\alpha0}}=\bra{\phi_{\beta0}}U\ket{\phi_{\beta0}}$ holds. 
With this decomposition, 
which equivalently maximizes Eq.~(\ref{eq:general.mix.d_kmax}) in terms of $\bm{\epsilon}$,
we can quantify $D$ as
\begin{align}
D
&=
|c_{\alpha}|^2\max_{\bm{w}_\alpha} \left[2 L(\bm{w}_\alpha,\pi_{\alpha0},\pi_{\alpha1}) -1 \right]
+|c_{\beta}|^2\max_{\bm{w}_\beta} \left[2 L(\bm{w}_\beta,\pi_{\beta0},\pi_{\beta1}) -1 \right]
\label{eq:general.mix.d_dcomp} \\
&= |c_{\alpha}|^2 \frac12 \Tr \left| \pi_{\alpha0}-\pi_{\alpha1} \right| 
+ |c_{\beta}|^2 \frac12 \Tr \left| \pi_{\beta0}-\pi_{\beta1} \right| 
\label{eq:general.mix.d_dcomp1} \\
&= |c_{\alpha}|^2 D_\alpha 
+ |c_{\beta}|^2 D_\beta
.
\label{eq:general.mix.d_dcomp2}
\end{align}
Note that $D_\alpha=D_\beta=D$ under the condition used in the decomposition.
In this way, we can quantify $D$ for \textcolor{black}{given} mixed states $\rho_0$ and $\rho_1=U\rho_0U^\dag$
without need for accessing E.

One might consider to evaluate $D$ by applying the mixed density operators of
$\rho_0$ (Eq.~(\ref{eq:general.mixwwm.rho0})) and $\rho_1$ (Eq.~(\ref{eq:general.mixwwm.rho1}))
to Eq.~(\ref{eq:general.opd2}) or (\ref{eq:general.d_kmax}), 
instead of pure density operators $\pi_0$ and $\pi_1$. 
It is nothing but the method \cite{zela2018, englert1999}  
to obtain the conventionally defined path distinguishability $D_\mathrm{c}$ 
for the mixed-state WWM;  
\begin{alignat}{1}
D_\mathrm{c}\coloneqq{}%
&\frac12\Tr \bigl| \rho_0 - \rho_1 \bigr| 
\label{eq:general.mix.opd2} \\
={}%
&\max_{\bm{w}} \left[2 L(\bm{w},\rho_0,\rho_1) -1 \right]
\label{eq:general.mix.dc_kmax}\\
={}%
&\max_{\bm{w}_{}} \left\{
|c_{\alpha}|^2 \left[2 L(\bm{w}_{},\pi_{\alpha0},\pi_{\alpha1}) -1 \right]
+|c_{\beta}|^2 \left[2 L(\bm{w}_{},\pi_{\beta0},\pi_{\beta1}) -1 \right]
\right\}. 
\label{eq:general.mix.d_rotd} 
\end{alignat}
Comparing Eq.~(\ref{eq:general.mix.d_rotd}) with Eq.~(\ref{eq:general.mix.d_kmax}),
we get
$D_\mathrm{c}\le D$ in general, 
and $D_\mathrm{c} = D$ only if 
$\rho_0$ is a pure state.
In Eq.~(\ref{eq:general.mix.d_rotd}),
the measurement operator set $\bm{w}$ is optimized for the given $\rho_0$ and $\rho_1$, 
without taking separate maximization for $\pi_{\alpha0,1}$ and $\pi_{\beta0,1}$.
Also, $D_\mathrm{c}$ is independent of the choice of the set, $\pi_{\alpha0}$ and $\pi_{\beta0}$, 
in the decomposition. 
In other words, 
$D_\mathrm{c}$ underestimates the 
\textcolor{black}{
WPI
}%
without using the entire information 
shared by E and W through the entanglement.   
On the other hand, as mentioned above,
$D$ reflects the full information shared by E and W,
and thus it \textcolor{black}{satisfies} the WPD equality Eq.~(\ref{eq:general.wpd}).
In the following, 
we demonstrate 
a practical example 
to obtain the amount of $D$ and $D_\mathrm{c}$ in mixed WWM states,
and to examine the WPD relation.

\subsection{\textcolor{black}{Application to experiment}}\label{sec:framework}

\begin{figure}[!t]
\centering\includegraphics[scale=0.5]{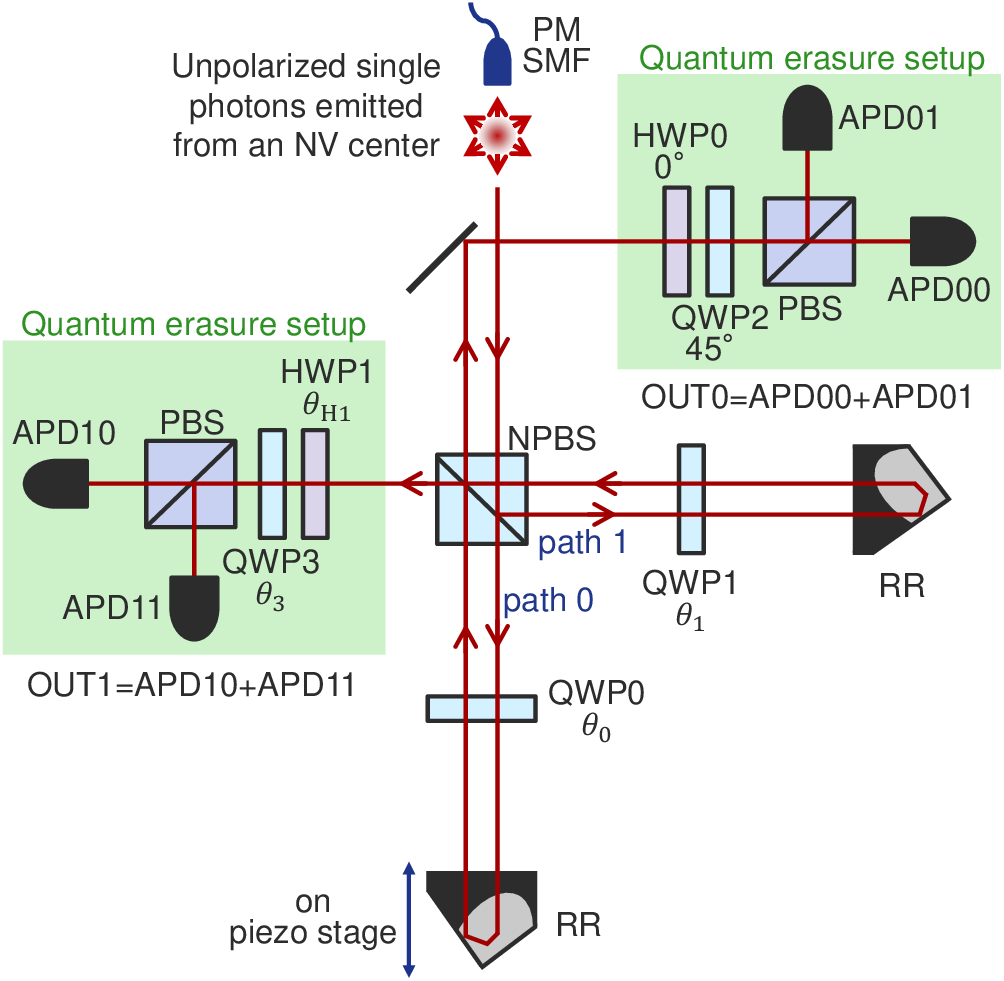}
\caption{\label{fig:setup} 
Michelson interferometer setup that we used 
\textcolor{black}{
in our experiments.
}%
PMSMF: 
a polarization-maintaining single-mode fiber, %
NPBS:  
a non-polarizing beamsplitter, %
RR:
retroreflectors, %
QWP(0\textcolor{black}{--}3): 
quarter-wave plates, %
HWP(0\textcolor{black}{--}1): 
half-wave plates, %
PBS: 
polarizing beamsplitters, %
APD(00\textcolor{black}{--}11): 
avalanche photodiodes. 
The green shaded areas represent 
\textcolor{black}{
the polarization-analyzing detectors
}%
\textcolor{black}{
used for the
}%
quantum erasure setup 
\textcolor{black}{
and the measurement of 
\textcolor{black}{the }%
path distinguishability. 
}%
 \textcolor{black}{
See the text for details.
}%
}
\end{figure}

We consider the Michelson interferometer shown in Fig.~\ref{fig:setup},
\textcolor{black}{
which sketches our experimental setup.
}%
The state of input photons can be described by 
\textcolor{black}{
$\rho_{\rm IN}\coloneqq
\rho_{\rm path} \otimes\rho_{\rm pol}= 
\dyad{0} \otimes\rho_{\rm pol}.$
}%
The first density matrix 
\textcolor{black}{
$\rho_{\rm path}$ 
}%
denotes
\textcolor{black}{
the 
}%
path qubit state 
\textcolor{black}{
of the photon
}%
with the 
bases
$\{\ket{0}, \ket{1}\}$, 
where $\ket{0}$ and $\ket{1}$ 
respectively 
\textcolor{black}{
represent
}%
the states
\textcolor{black}{
in which the photon passes
}%
through path 0 and path 1. 
The second density matrix $\rho_{\rm pol}$ denotes
\textcolor{black}{
the 
}%
polarization qubit state
\textcolor{black}{
of the photon
}%
with the 
bases 
$\{\ket{\rm H}, \ket{\rm V}\}$, 
where $\ket{\rm H}$ and $\ket{\rm V}$
respectively 
\textcolor{black}{
represent
}%
the horizontal and 
vertical polarization states. 
\textcolor{black}{
We use
}%
the polarization qubit state 
as 
\textcolor{black}{
the WWM.
}%

First, we derive 
\textcolor{black}{
the
}%
unitary matrix 
\textcolor{black}{
that describes 
}%
the transformation of the photon state in 
\textcolor{black}{
the
}%
Michelson interferometer. 
\textcolor{black}{
The
}%
unitary matrices representing the operation of a non-polarizing beamsplitter (NPBS) for 
\textcolor{black}{
horizontal (H)
}%
and 
\textcolor{black}{
vertical (V)
}%
polarizations can be respectively expressed as 
\textcolor{black}{
$U_{\rm H}\coloneqq (\sigma_0+i\sigma_1)/\sqrt{2}$ and $U_{\rm V}\coloneqq (\sigma_0-i\sigma_1)/\sqrt{2}$,
}%
where $\sigma_{0}$ is the two-dimensional identity matrix and $\sigma_{k=1, 2, 3}$ are 
\textcolor{black}{
the
}%
Pauli %
\textcolor{black}{%
matrices. 
}%
Note that 
the off-diagonal elements 
of 
\textcolor{black}{
$U_{\rm H}$ and $U_{\rm V}$
have opposite signs, 
}%
according to Fresnel's definition of coordinates for reflection/refraction 
\textcolor{black}{
of p- (H in our configuration) and s- (V) polarized light.
}%
Thus, 
\textcolor{black}{
the
}%
total unitary matrix of 
\textcolor{black}{
an
}%
NPBS is expressed as
\begin{equation}
U_{\rm BS}\coloneqq 
\textcolor{black}{
U_{\rm H}
}%
\otimes\dyad{\rm H}+
\textcolor{black}{
U_{\rm V}
}%
\otimes\dyad{\rm V}=
\frac{1}{\sqrt{2}}
\begin{pmatrix}
\begin{array}{cc}
\sigma_{0} & i\sigma_{3}  \\
i\sigma_{3} & \sigma_{0}
\end{array}
\end{pmatrix}
. 
\label{eq:BS}
\end{equation}
The 
\textcolor{black}{
polarization transformation matrices 
}%
(Jones matrices) of a mirror, a quarter-wave plate (QWP), and a half-wave plate (HWP) 
with the bases 
$\{\ket{\rm H}, \ket{\rm V}\}$ are respectively defined by %
$M\coloneqq \sigma_{3}, U_{\rm QWP}(\theta)\coloneqq
\frac{i}{\sqrt{2}}[-i\sigma_0+\sin2\theta\sigma_1+\cos2\theta\sigma_3], 
U_{\rm HWP}(\theta)\coloneqq
\sin2\theta\sigma_1+\cos2\theta\sigma_3$, 
where $\theta$ is 
\textcolor{black}{
the
}%
angle of the fast axis 
\textcolor{black}{
(measured from the horizontal axis; the same shall apply hereafter)
}%
of the wave plate.
\textcolor{black}{
In our interferometer,
}%
we used 
a couple of 
polarization rotators each of which consists of a double-pass QWP and a retroreflector; 
its unitary matrix $U_{\rm R}$ is expressed by 
\begin{equation}
U_{\rm R}(\theta) \coloneqq
U_{\rm QWP}(-\theta)MU_{\rm QWP}(\theta)
=ie^{4i\theta\sigma_{2}}.
\label{eq:ur}
\end{equation}
\textcolor{black}{
Note that
}%
$U_{\rm R}(\theta)$
is equivalent to polarization rotation 
\textcolor{black}{
given by
}%
a half-wave plate followed by a mirror,
i.e., $M U_{\rm HWP}(\theta)$,  
and the Jones matrix of the three-times reflection on a retroreflector becomes that of a single mirror
($M^3=M$).
\textcolor{black}{
We added WPI by applying different polarization rotations between path 0 and 1.
}%
The transformation unitary matrix 
through 
\textcolor{black}{
both paths
including the 
}%
QWPs and the retroreflectors 
is represented by 
\begin{equation}
U_{\rm W}\coloneqq 
e^{i\phi}\dyad{0}\otimes U_{\rm R}(\theta_{0})+
\dyad{1}\otimes U_{\rm R}(\theta_{1}), 
\label{eq:UOUT}
\end{equation}
where 
$\phi$ is the 
phase 
\textcolor{black}{
change added
}%
to path 0; 
$\theta_{0}$ and $\theta_{1}$ are respectively the fast axis angles of QWP0 and QWP1. 
\textcolor{black}{
Maximum WPI is added when, e.g., 
$\theta_{0}=0$ and $\theta_{1}=\pi/4$.
}%
\textcolor{black}{
The total
}%
transformation unitary matrix of this interferometer 
becomes
$U_{\rm I}\coloneqq U_{\rm BS}^\dagger U_{\rm W}U_{\rm BS}$. 
Thus, we can obtain the polarization density matrix of 
\textcolor{black}{
the output of
}%
path 1 by 
\begin{alignat}{1}
\rho_{\rm pol}^{\rm OUT1}\coloneqq{}%
&\Tr_{\rm path} [\{(\dyad{1}\otimes \textcolor{black}{\sigma_0})U_{\rm I}\}\rho_{\rm IN}\{(\dyad{1}\otimes \textcolor{black}{\sigma_0})U_{\rm I}\}^\dagger]\\
={}%
&\bra{1}U_{\rm I}\rho_{\rm IN}U_{\rm I}^\dagger\ket{1}
\end{alignat}
and the existing probability of a photon at the output of path 1 by
\begin{equation}
I_1\coloneqq \Tr \rho_{\rm pol}^{\rm OUT1}=\frac{1}{2}[1+V \cos(\phi+\arg \textcolor{black}{C_\mathrm{i}})], 
\label{eq:I1}
\end{equation}
where 
\begin{equation}
\textcolor{black}{C_\mathrm{i}}\coloneqq \Tr [U_{\rm R}(\theta_{0})\rho_{\rm pol}\tilde{U}_{\rm R}^\dagger(\theta_{1})]
=\left(\Tr [
\textcolor{black}{U_\mathrm{S}}
\rho_{\rm pol}]\right)^*, 
\label{eq:cdef}
\end{equation} 
$\tilde{U}_{\rm R}\coloneqq \sigma_3 U_{\rm R}\sigma_3$, 
\textcolor{black}{$U_\mathrm{S}\coloneqq U_{\rm R}^\dagger(\theta_{0})\tilde{U}_{\rm R}(\theta_{1})$,} %
and the visibility $V$ becomes 
\begin{equation}
V\coloneqq |C_\mathrm{i}|=\left|\Tr [
\textcolor{black}{
U_\mathrm{S}
}%
\rho_{\rm pol}]\right|. 
\end{equation}

Next, we 
\textcolor{black}{examine}
the conventional  path distinguishability $D_\mathrm{c}$ in this setup.
\textcolor{black}{
In order to quantify the distinguishability of the polarization states passing through either path~0 or path~1 of the interferometer,
we} consider the polarization density matrix 
$\rho_{{\rm pol}p}^{\rm OUT1}$ for the output of path~1 
when 
path $p$ 
in the interferometer
is open while the other path is blocked: 
\begin{equation}
\rho_{{\rm pol}p}^{\rm OUT1}\coloneqq \bra{1}\Pi_{p}\rho_{\rm IN}\Pi_{p}^\dagger\ket{1}
\label{eq:rout1}, 
\end{equation}
where 
$\Pi_{p}\coloneqq U_{\rm BS}^\dagger [\dyad{p}\otimes \sigma_0]U_{\rm W}U_{\rm BS}$.
The conventional
distinguishability $D_\mathrm{c}$
\textcolor{black}{given in Eq.~(\ref{eq:general.mix.opd2})
leads}
\begin{alignat}{1}
D_\mathrm{c}
={}%
&\frac{1}{2}\Tr \left| \rho_{\rm pol0}^{\rm OUT1}-\rho_{\rm pol1}^{\rm OUT1}\right|
\label{eq:d_cond}\\
={}%
&\frac{1}{2}\Tr \left| U_{\rm R}(\theta_{0})\rho_{\rm pol}U_{\rm R}^\dagger(\theta_{0})-
\tilde{U}_{\rm R}(\theta_{1})\rho_{\rm pol}\tilde{U}_{\rm R}^\dagger(\theta_{1}) \right|
\\
\textcolor{black}{=}{}%
&\textcolor{black}{\frac{1}{2}\Tr 
\left|\rho_{\rm pol} -
U_\mathrm{S}\rho_{\rm pol}U_\mathrm{S}^\dagger \right|
}%
.
\label{eq:d_rotd} 
\end{alignat}

In the following, by representing the rotated density matrices in Eq.~(\ref{eq:d_rotd}) using Stokes parameters according to \cite{zela2018}, we obtain simpler and intuitive representations for $V$ and $D_\mathrm{c}$. 
Let the input polarization density matrix $\rho_{\rm pol}$ be represented using Stokes parameters as 
$\rho_{\rm pol}=\frac{1}{2}(\sigma_0+ \bm{s}\cdot \bm{\sigma})$, 
where $\bm{s}\coloneqq(s_1, s_2, s_3)$ is
\textcolor{black}{
the Stokes vector
$(0\le|\bm{s}|\le1)$ 
}%
and $\bm{\sigma}\coloneqq (\sigma_{1}, \sigma_{2}, \sigma_{3})$. 
Letting 
$\bm{s}_0$ 
and $\bm{s}_1$ be
respectively 
the corresponding Stokes vectors of $\rho_{\rm pol0}^{\rm OUT1}$ and $\rho_{\rm pol1}^{\rm OUT1}$, 
we can write $\bm{s}_0=R_0\bm{s}$ and $\bm{s}_1=R_1\bm{s}$, where $R_{0}$ and $R_{1}$ are rotation matrices of the Stokes 
\textcolor{black}{
vector
}%
associated with $U_{\rm R}(\theta_{0})$ and $\tilde{U}_{\rm R}(\theta_{1})$,
\textcolor{black}{
respectively.
For instance, we see from Eq.~(\ref{eq:ur}) that 
$U_{\rm R}(\theta)$ in our case corresponds to $4\theta$ rotation of the Stokes vector about $(0,1,0)$ axis. 
}%
\textcolor{black}{
Using 
the Stokes vector representation,
}%
$D_\mathrm{c}$ is given by  
\begin{equation}
D_\mathrm{c}
=\frac{1}{2}\left| \bm{s}_0-\bm{s}_1\right|
\textcolor{black}{
=\frac{1}{2}\left|\bm{s}-R_\mathrm{S}\bm{s}\right|, 
}%
\label{eq:Dc-s}
\end{equation}
\textcolor{black}{
i.e.,
}%
half 
the Euclidean distance between $\bm{s}_0$ and $\bm{s}_1$,
\textcolor{black}{
where
$R_\mathrm{S} = R_0^{-1}R_1$.
}%
\textcolor{black}{
In general,
}%
the rotation matrix 
$R_\mathrm{S}$ 
associated with 
\textcolor{black}{
$U_\mathrm{S}$
}%
can be written as a rotation of angle 
\textcolor{black}{
$\Omega$ 
}%
about an axis $\hat{\bm{n}}$,
\textcolor{black}{
i.e.,
$
U_\mathrm{S}
=e^{-i\Omega\hat{\bm{n}}\cdot\bm{\sigma}}
= e_0\sigma_{0}+i\bm{e}\cdot \bm{\sigma}
$, 
where 
$e_0\coloneqq\cos(\Omega/2)$ and  $\bm{e}\coloneqq\sin(\Omega/2)\hat{\bm{n}}$
.
}%
Then, using the Euler-Rodrigues formula, we can write \cite{zela2018}
\begin{alignat}{2}
&V%
&&=\sqrt{e_0^2+(\bm{e}\cdot\bm{s})^2}, 
\label{eq:vs}\\ 
&D_\mathrm{c}%
&&=\sqrt{\bm{e}^2\bm{s}^2-(\bm{e}\cdot\bm{s})^2}, 
\label{eq:ds} 
\end{alignat}
\textcolor{black}{
and
}%
hence, 
\begin{equation}
V^2+
D^2_\mathrm{c}
=e_0^2+\bm{e}^2\bm{s}^2=
\textcolor{black}{
\cos^2\left(\frac{\Omega}{2}\right)+\bm{s}^2\sin^2\left(\frac{\Omega}{2}\right). 
}%
\label{eq:wpdcon}
\end{equation}
\textcolor{black}{
From Eq.~(\ref{eq:wpdcon}) and $|\bm{s}|\le1$, 
we see that $0\leq V^2+D^2_\mathrm{c}\leq1$. 
}%
\textcolor{black}{
The WPD inequality
\textcolor{black}{
(Eq.~(\ref{eq:wpd}))
}%
in terms of $V$ and $D_\mathrm{c}$ is thus derived.
Note that, 
when the input polarization state is in a pure state ($|\bm{s}|=1$),
the WPD inequality is saturated, i.e., $V^2+D^2_\mathrm{c}=1$. 
}%
When
the input polarization state is in a mixed state
($|\bm{s}|<1$), however,
Eq.~(\ref{eq:wpdcon}) gives
$0\leq V^2+D^2_\mathrm{c}<1$
unless 
\textcolor{black}{
$\Omega=0$.
}%
In particular,
when the input polarization state is in the completely mixed state ($\bm{s}=\bm{0}$), 
\textcolor{black}{
it turns out that
$D_\mathrm{c}=0$ 
}%
whereas 
\textcolor{black}{
$0\le V^2=\cos^2(\Omega/2)\le1$,
}%
implying
that the conventional $D_\mathrm{c}$ does not adequately quantify the amount of WPI, which on the other hand decreases $V$ even in mixed states. 

\textcolor{black}{
Next, 
we consider quantifying $D$ in terms of Eq.~(\ref{eq:general.mix.d_dcomp}) or (\ref{eq:general.mix.d_dcomp1}).
}%
Any mixed state can be decomposed as a convex combination or a weighted sum of two pure polarization states, regardless of what states were actually prepared with. 
Let 
\textcolor{black}{
$\rho_{\rm pol}=p_{\alpha} \pi_{\alpha}+p_{\beta} \pi_{\beta}$ 
}%
be the decomposition of an input polarization state 
$\rho_{\rm pol}$
into
a convex combination of 
two pure polarization states
\textcolor{black}{
$\pi_\alpha=\ketbra{\phi_\alpha}$ and 
$\pi_\beta=\ketbra{\phi_\beta}$,
}%
where 
\textcolor{black}{
$p_{\alpha,\beta}\geq 0$ and $p_{\alpha}+p_{\beta}=1$.
}%
Accordingly, the Stokes vector 
$\bm{s}$ corresponding to $\rho_{\rm pol}$
is also decomposed as
\textcolor{black}{
$\bm{s}=p_{\alpha}\hat{\bm{s}}_{\alpha}+p_{\beta}\hat{\bm{s}}_{\beta}$,
}%
where
\textcolor{black}{
$\hat{\bm{s}}_{\alpha,\beta}$
}%
are the Stokes vectors corresponding to 
\textcolor{black}{
$\pi_{\alpha,\beta}$, 
}%
respectively. 
Note that
$\bm{s}$ 
thus corresponds to an
interior point of 
\textcolor{black}{
$\hat{\bm{s}}_{\beta}-\hat{\bm{s}}_{\alpha}$, 
}%
and
that 
\textcolor{black}{
$\pi_{\beta}$ ($\hat{\bm{s}}_{\beta}$) 
}%
and 
\textcolor{black}{
$p_{\alpha,\beta}$ 
}%
are uniquely determined when 
$\rho_{\rm pol}$ ($\bm{s}$) 
and 
\textcolor{black}{
$\pi_{\alpha}$ ($\hat{\bm{s}}_{\alpha}$) 
}%
are given. 
\textcolor{black}{
The condition for appropriate decomposition in which 
Eq.~(\ref{eq:general.mix.d_dcomp}) or (\ref{eq:general.mix.d_dcomp1})
gives correct $D$ is that 
$\bra{\phi_{\alpha}}U_{01}\ket{\phi_{\alpha}}=\bra{\phi_{\beta}}U_{01}\ket{\phi_{\beta}}$.
}%
We find that 
such
decomposed pure states 
are 
the
pure states that satisfy 
\textcolor{black}{
$\hat{\bm{n}}\cdot\hat{\bm{s}}_{\alpha}=\hat{\bm{n}}\cdot\hat{\bm{s}}_{\beta}=\hat{\bm{n}}\cdot\bm{s}$%
}%
. 
For these states, 
we obtain
\begin{equation}
D=\sqrt{\bm{e}^2-(\bm{e}\cdot\bm{s})^2}. %
\label{eq:d_redefs2}
\end{equation}
With Eq.~(\ref{eq:vs}), we get
$V^2+D^2 = 1$, i.e., 
Eq.~(\ref{eq:general.wpd}).
Thus, 
\textcolor{black}{
we see that
}%
the WPD equality 
Eq.~(\ref{eq:general.wpd})
is always satisfied for both pure and mixed input states. %
\textcolor{black}{
At this point, 
it is worth examining 
the relation between $D$ and $D_\mathrm{c}$ in terms of entanglement.
From Eqs.~(\ref{eq:ds}) and (\ref{eq:d_redefs2}),
we obtain 
\begin{equation}
D^2
=D_\mathrm{c}^2 + (1-\bm{s}^2)\bm{e}^2
=D_\mathrm{c}^2 + C_\mathrm{WE}^2\bm{e}^2,
\label{eq:D-D_C-e}
\end{equation}
where
\textcolor{black}{
$C_\mathrm{WE}=\sqrt{1-\bm{s}^2}=\sqrt{2(1-\gamma_\mathrm{pol})}$
}%
is the concurrence of the mixed-state WWM (W) entangled with the external system (E),
as shown in Eq.~(\ref{eq:general.mixwwm}),
\textcolor{black}{
and $\gamma_\mathrm{pol}=\Tr\rho_\mathrm{pol}^2$ is the purity of the input polarization state $\rho_\mathrm{pol}$.
}%
The last term in Eq.~(\ref{eq:D-D_C-e}) represents the portion of distinguishability 
that is not taken into account in $D_\mathrm{c}$.
This `hidden' portion of distinguishability
is shared between W and E through the entanglement,
and can be in principle recovered by accessing not only W but also E.
Instead, as shown above, it can also be recovered by optimized decomposition of the mixed-state WWM into two pure states.
In this case, 
the recovery of the hidden distinguishability is
mediated by
the distinguishability (i.e., incoherence) between the decomposed pure states,
which complementarily results in the degradation of visibility
when polarization rotation between paths takes place.
Thus, we understand that 
$D$ incorporates the distinguishability not only held in the mixed-state WWM but also shared through the entanglement,
leading to the WPD equality, Eq.~(\ref{eq:general.wpd}). 
}%

\begin{figure}[!t]
	\centering\includegraphics[width=85mm, keepaspectratio, clip]{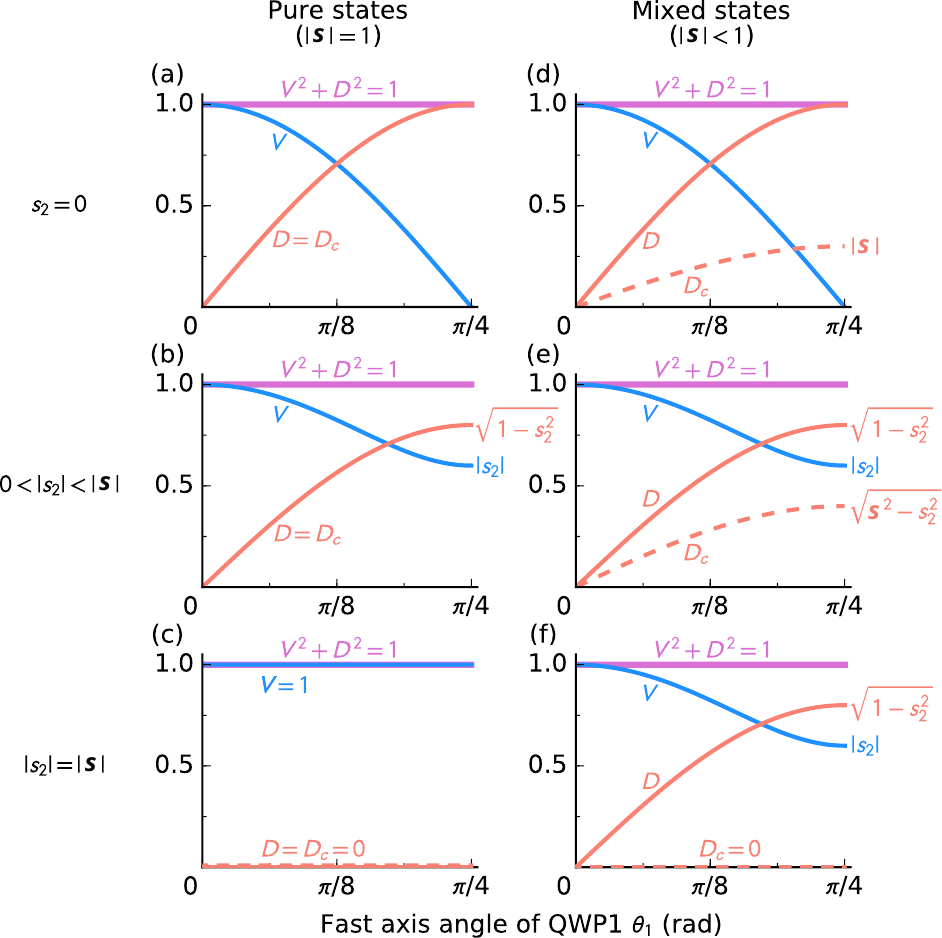}%
	%
	%
	%
	\caption{\label{fig:wpd6}
		The 
visibility $V$, the conventional path distinguishability $D_\mathrm{c}$, the redefined path distinguishability $D$, and $V^2+D^2$		
are 
depicted as 
functions
of 
$\theta_1$
(the fast axis angle of 
QWP1 in Fig.~\ref{fig:setup}), 
which varies 
\textcolor{black}{%
the
}%
WPI. 
Their behaviors 
are classified into six cases,
\textcolor{black}{%
(a)%
\textcolor{black}{%
--%
}%
(f), as described in the text.
}%
}
\end{figure}

Now we examine the 
\textcolor{black}{
amounts of $V$, $D_\mathrm{c}$, and $D$ as well as the WPD relation in the
}%
case of our experimental setup depicted in Fig.~\ref{fig:setup}.
Using 
Eq.~(\ref{eq:ur})
and
$\textcolor{black}{
U_\mathrm{S}
}%
=e^{-4i(\theta_0+\theta_1)\sigma_{2}}$, 
we obtain  
\textcolor{black}{
$\Omega=-4\theta_1$, 
}%
$e_0=\cos2\theta_1$, 
\textcolor{black}{
and
}%
$\bm{e}=\sin2\theta_1(0,-1,0)$ for $\theta_0=0$. 
\textcolor{black}{Equations~(\ref{eq:vs})--(\ref{eq:d_redefs2})}
turn out to be
\begin{alignat}{3}
&&&V{}&&=\sqrt{\cos^2 2\theta_1+s_2^2\sin^2 2\theta_1}, 
\label{eq:vnews}\\
&&&D_\mathrm{c}{}&&=\sqrt{\bm{s}^2-s_2^2}\ \textcolor{black}{\left|\sin2\theta_1\right|} 
\label{eq:dcons}
\textcolor{black}{
, 
}%
\\
&&&D{}&&=\sqrt{1-s_2^2}\ \textcolor{black}{\left|\sin2\theta_1\right|}, 
\label{eq:dnews}\\
&V^2+{}&&D^2_\mathrm{c}&&=\cos^2 2\theta_1+\bm{s}^2\sin^2 2\theta_1%
\textcolor{black}{
, 
}%
\label{eq:wpdcons}\\
&V^2+{}&&D^2&&=1, 
\label{eq:wpdnews}
\end{alignat}
where $s_2$ is the Stokes parameter representing the degree of circular polarization of the input photon. 
The behavior of the visibility $V$ and the conventional distinguishability 
$D_\mathrm{c}$ %
in this case can be classified into six cases by pure or mixed states and values of $s_2$, e.g., 
pure states ($|\bm{s}|=1$) with $s_2=0$ (a) and $0<|s_2|<1$ (b), 
pure circular polarizations $|s_2|=1$ (c), 
mixed states ($0\leq|\bm{s}|<1$) 
with $s_2=0$ (d), 
\textcolor{black}{
$0<|s_2|<|\bm{s}|$ 
}%
(e), 
and 
\textcolor{black}{
$|s_2|=|\bm{s}|$ 
}%
(f) 
as shown in Fig.~\ref{fig:wpd6} 
\textcolor{black}{
together with
}%
the behavior of
the redefined path distinguishability $D$. 
\textcolor{black}{
Note that 
the completely unpolarized state ($\bm{s}=\bm{0}$) is
included in the cases (d) and (f). 
}%
\textcolor{black}{
As described earlier, 
$D$ exhibits the same behavior regardless of whether the state is pure or mixed. 
}%
\textcolor{black}{
In our case, 
$V$, $D_\mathrm{c}$, and $D$ 
}%
depend only on $s_2$ among the three Stokes 
\textcolor{black}{
vector components 
}%
because
\textcolor{black}{
the 
}%
polarization rotation (Eq.~(\ref{eq:ur}))
about the 
\textcolor{black}{
$(0,1,0)$ axis is used 
}%
to add WPI.
\textcolor{black}{
}%
In general, 
WPI is embedded in the Stokes vector components 
normal to the rotation axis $\hat{\bm{n}}$ and %
accordingly 
$D$ and $D_\mathrm{c}$ are proportional to
$\sqrt{1-s_n^2}$ 
for pure states, 
where $s_n$ is the Stokes vector component parallel to $\hat{\bm{n}}$. 
In this case, 
the amount of $D^2$ (or $D_\mathrm{c}^2$) and $V^2$ complement each other so that 
the perfect complementary relation 
$V^2+D^2=V^2+D_\mathrm{c}^2=1$ 
is satisfied.
However, in mixed states, $D_\mathrm{c}$ is proportional to
$\sqrt{\bm{s}^2-s_n^2}$ %
($|\bm{s}|<1$), 
which 
takes only the coherent portion (proportional to $|\bm{s}|$) into account
and thus the complementation is incomplete.
$D$ is defined, using   
the decomposition of a mixed state 
into
a convex combination of two pure polarization states that satisfy
\textcolor{black}{%
$\hat{\bm{n}}\cdot\hat{\bm{s}}_{\alpha}=\hat{\bm{n}}\cdot\hat{\bm{s}}_{\beta}=\hat{\bm{n}}\cdot\bm{s}$,
}%
so that it again becomes proportional to 
$\sqrt{1-s_n^2}$, 
as if each photon has the pure polarization state that satisfies the relation above.
Note also that, as stated earlier,
$D$ by definition implies the 
distinguishability 
between
the decomposed pure states. 

\section{Experimental methods}
The experimental setup is shown in Fig.~\ref{fig:setup}. 
As a photon source, we used unpolarized \textcolor{black}{true} single photons emitted from a [111]-oriented NV center in (111) diamond \cite{abe2017}%
\textcolor{black}{%
, which exhibit the anti-bunching property}. 
This source naturally emits unpolarized single photons owing to thermal mixing of the electronic excited states responsible for two orthogonally polarized emissions of the NV center. 
\textcolor{black}{
The use of true single photons ensures that each photon is solely and individually injected into the interferometer 
and thus allows rigorous verification of the WPD that belongs to a single quantum object \cite{englert1996, englert1999}, 
although the measures ($V$, $D_\mathrm{c}$, $D$, etc.) of our experiment are independent of photon statistics, or even they can be measured using classical light.} %
\textcolor{black}{
The emission from a single NV center
was filtered spectrally by a band-pass filter
}%
(center wavelength: 679~nm, bandwidth: 36~nm)
\textcolor{black}{
and spatially by a polarization-maintaining single-mode fiber.  
}%
The single-photon nature 
\textcolor{black}{
was evaluated by
the second-order correlation function 
$g^{(2)}(\tau)$ 
between photons at intervals of $\tau$.
}%
We obtained
\textcolor{black}{
$g^{(2)}(0)=0.18$;
}%
\textcolor{black}{
}%
\textcolor{black}{
the anti-bunching property with 
$g^{(2)}(0)<0.5$
demonstrated the apparent
}%
single-photon nature of the source. 
The degree of static 
\textcolor{black}{
polarization
}%
was evaluated by 
state tomography 
\cite{peters2003}. 
The estimated Stokes vector 
was 
$
(s_1, s_2, s_3)
= 
(-0.003(9), -0.004(15), 0.061(6))$; 
the corresponding degree of polarization was 
0.0065(9)
and 
the fidelity 
to the completely unpolarized state 
was
0.9994(2).
\textcolor{black}{
Note that the values inside the parentheses denote 68.3\% confidence intervals.
}%
The 
dynamical 
\textcolor{black}{
statistics in polarizations between photons
}%
was evaluated by the polarization correlation function
defined by 
$C_\mathrm{P}(\tau)\coloneqq
(g^{(2)}_\mathrm{HH}(\tau)+g^{(2)}_\mathrm{VV}(\tau)-g^{(2)}_\mathrm{HV}(\tau)-g^{(2)}_\mathrm{VH}(\tau))/\sum_{i,j\in{\mathrm{H}, \mathrm{V}}} g^{(2)}_{ij}(\tau)$, 
where $g^{(2)}_{ij}(\tau)$ indicates 
\textcolor{black}{
the second-order correlation function
}%
$g^{(2)}(\tau)$ 
\textcolor{black}{
between photons
in
}%
$i-$ and $j-$polarizations 
\cite{abe2017}.   
We found that 
\textcolor{black}{
$C_\mathrm{P}(\tau)$ 
was close to zero regardless of $\tau$;
the mean and the standard deviation of $C_\mathrm{P}(|\tau|\le 200~\mathrm{ns})$ 
were -0.003 and 0.006, respectively. 
}%
These results 
\textcolor{black}{
demonstrate that our source has 
}%
almost ideal %
properties of unpolarized single photons 
\textcolor{black}{
not only in statical statistics expressed by the density matrix (Stokes vector)
but also in dynamical randomness between photons expressed by the correlation function\cite{abe2017}. 
}%
\textcolor{black}{
The
}%
unpolarized photons 
\textcolor{black}{
thus generated
}%
were fed into the Michelson interferometer
\textcolor{black}{
depicted in Fig.~\ref{fig:setup}.
As explained earlier in Sec.~\ref{Sec.Theory},
WPD of the unpolarized single photons is examined 
by adding WPI in the polarization degrees of freedom,
}%
using two quarter-wave plates
(QWP0 and QWP1) in 
\textcolor{black}{
path 0 and path 1
}%
of the interferometer. 
\textcolor{black}{
Photons outputted from the interferometer
to the output path 0 (path 1) 
}%
were detected 
\textcolor{black}{
by
a pair of
}%
avalanche photodiodes
\textcolor{black}{
 APD00 and APD01
 (APD10 and APD11),
which are
}%
designed to detect photons 
on 
\textcolor{black}{
an arbitrary pair of orthogonal 
}%
polarization 
\textcolor{black}{
bases
}%
\textcolor{black}{
using half-wave plates (HWP0 and HWP1), 
quarter-wave plates (QWP2 and QWP3), and polarizing beamsplitters (PBS).
\textcolor{black}{
Detection on circular-polarization bases
}%
is used to carry out the quantum erasure experiment.
}%
\textcolor{black}{
Detection on arbitrary linear-polarization bases
is used to evaluate the conventional distinguishability $D_\mathrm{c}$ for linear polarization inputs,
	as described in Sec.~4.3.
}%
\textcolor{black}{
Unpolarized detection, i.e., OUT0 (OUT1) is carried out
by summing the detection events 
of APD00 and APD01 (APD10 and APD11)
for path 0 (path 1).
}%

\section{Experimental results and discussion}
\subsection{\textcolor{black}{Bare coherence properties of the photon source}}
\begin{figure*}[!t]
    \centering\includegraphics[scale=1]{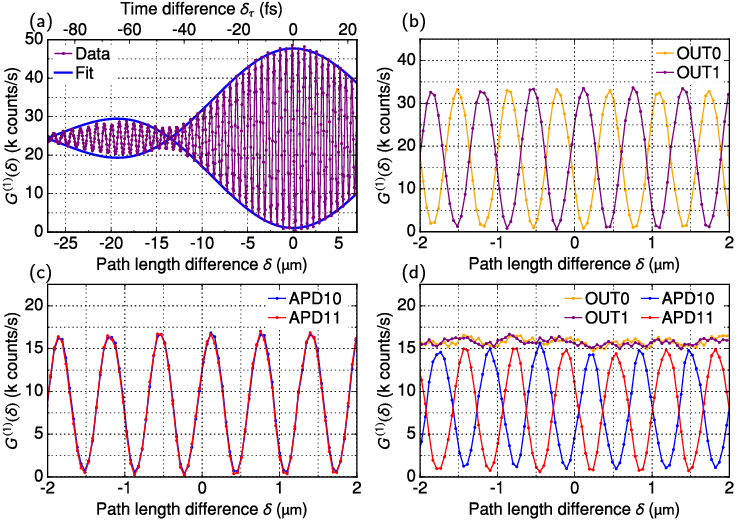}
	\caption{\label{fig:g1} 
Interference 
\textcolor{black}{
fringes ($G^{(1)}$) observed in the experiments using
}%
unpolarized single photons. 
(a) 
$G^{(1)}$ 
\textcolor{black}{
observed at 
}%
OUT1 
without QWP0 and QWP1.
(b) 
$G^{(1)}$ 
\textcolor{black}{
observed at
}%
OUT0 (yellow) and OUT1 (purple) when 
WPI is completely absent 
\textcolor{black}{
($\theta_0=\theta_1=0^\circ$).
}%
(c) 
\textcolor{black}{
Same as (b) but
}%
\textcolor{black}{
observed at
}%
APD10 (blue) and APD11 (red),
which 
\textcolor{black}{
detect photons of OUT1  
}%
on the circular polarization 
\textcolor{black}{
bases. 
}%
(d)
$G^{(1)}$ 
\textcolor{black}{
observed at
OUT0 (yellow), OUT1 (purple),
APD10 (blue), and APD11 (red)
}%
when 
\textcolor{black}{
maximum
}%
WPI is 
added 
\textcolor{black}{
($\theta_0=0^\circ$ and $\theta_1=45^\circ$).
}%
}
\end{figure*}

We measured the 
(unnormalized) 
first-order autocorrelation function $G^{(1)}$ \cite{hertel2015, jelezko2003, braig2003, marshall2011, jacques2005, jacques2007} 
\textcolor{black}{
of the photon source by observing the interference fringes of 
}%
OUT1 in Fig.~\ref{fig:setup}
without 
QWP0 and QWP1, 
\textcolor{black}{
using a single APD instead of
}%
the quantum erasure setup
(green-shaded areas shown in Fig.~\ref{fig:setup}).
The result is shown in Fig.~\ref{fig:g1}(a). 
\textcolor{black}{
The data exhibit
}%
a high visibility $V$ of 
\textcolor{black}{
0.983,
demonstrating intrinsic high coherence of our source.
}%
\textcolor{black}{
The envelope of the fringes was fitted by the function
}%
$A(1+V{\rm sinc}[(\delta-\delta_0)/l_\mathrm{c}])$, 
where 
$A$ is the base
\textcolor{black}{
level,
}%
$V$ the visibility,
$\delta$ 
the path length difference, 
$\delta_0$ 
the position of 
zero path 
\textcolor{black}{
length
}%
difference,
and
$l_\mathrm{c}$ the coherence length
\cite{hertel2015}. 
\textcolor{black}{
From the fit, we obtain 
$V=0.956$ and $l_\mathrm{c}=13.5~\mu\text{m}$.
}%
Note that the sinc-function-like shape of the envelope of $G^{(1)}$ 
\textcolor{black}{
with $l_\mathrm{c}=13.5~\mu\text{m}$
is consistent with the almost
}%
rectangular 
\textcolor{black}{
shape
}%
of the emission spectrum
\textcolor{black}{
resulting from the 
}%
band-pass filtering 
\textcolor{black}{
(center wavelength: 679~nm, bandwidth: 36~nm).
}%
\textcolor{black}{
The slight decrease in $V$ obtained from the fit is in part caused by the smaller amplitude of the second peak (around $\delta=-20~\mu$m) 
of the data than that expected from the exact sinc function.
}%

\subsection{\textcolor{black}{Interference with WPI and quantum erasure}}
\textcolor{black}{
Here we demonstrate
}%
WPI-induced destruction and quantum-erasure-induced revival \cite{kim2000, walborn2002, gogo2005} of interference
\textcolor{black}{
in the case of
}%
unpolarized single photons. 
We measured $G^{(1)}$ under extreme WPI conditions,
\textcolor{black}{
i.e., 
no WPI ($\theta_0=\theta_1=0^\circ$ in Fig.~\ref{fig:setup})
and 
the maximum WPI ($\theta_0=0^\circ$ and $\theta_1=45^\circ$).
}%
\textcolor{black}{
The effect of WPI on 
the completely unpolarized photons 
and the properties of quantum erasure are 
discussed in Sec.~\ref{Sec.Theory} and {Supplement 1}. 
In short,
unpolarized photons  
exhibit vanishing interference fringes
under the maximum WPI,
even though 
the polarization density matrix
is identical and unchanged in both paths.
The WPI in this case is printed in the path coherence
between orthogonal polarization modes in path 0 and path 1.
If we observe the output in a circular polarization basis,
the interference fringes revive with high visibility (quantum erasure).
}%

The observed interference fringes for completely unpolarized photons in the case of no WPI
are shown in Figs.~\ref{fig:g1}(b) and (c). 
Figure~\ref{fig:g1}(b) shows 
$G^{(1)}$ 
\textcolor{black}{
measured at
}%
OUT0 (yellow) 
and OUT1 (purple)%
\textcolor{black}{
; they
}%
exhibit complementary interference fringes with high 
\textcolor{black}{
visibilities 
}%
\textcolor{black}{
($V=0.953$ and 0.966, respectively).  
}%
Figure~\ref{fig:g1}(c) shows $G^{(1)}$ 
\textcolor{black}{
measured at 
}%
APD10 (blue) and APD11 (red),
both of which show 
in-phase
interference fringes 
\textcolor{black}{
($V=0.956$ and 0.977, respectively).  
}%
\textcolor{black}{
Thus, in the case of no WPI,
APD10 and APD11 exhibit almost perfect, in-phase interference fringes,
which result in the perfect interference in OUT1 (and OUT0).
}%

The observed $G^{(1)}$ signals in the case of the maximum WPI
are shown in Fig.~\ref{fig:g1}(d). 
We see that 
the interference fringes
\textcolor{black}{
in
}%
OUT0 and OUT1 almost disappear
with vanishing 
\textcolor{black}{
visibilities 
}%
\textcolor{black}{
($V=0.076$ and 0.065, respectively). 
}%
\textcolor{black}{
Thus, we confirm that
}%
the maximum WPI 
\textcolor{black}{
is 
}%
added to the completely unpolarized state.
Nevertheless,
if we observe the signals split 
\textcolor{black}{
into
}%
APD10 (blue) and APD11 (red), 
interference fringes revive 
with high 
\textcolor{black}{
visibilities ($V=0.885$ and 0.923, respectively),
}%
as a consequence of the quantum erasure protocol.
The interference fringes 
\textcolor{black}{
observed in
}%
APD10 and APD11 are complementary, i.e., out of phase,  
\textcolor{black}{
to
}%
each other,
\textcolor{black}{
resulting in the vanishing interference fringes in OUT1 (and OUT0).
}%
These results evidently demonstrate the realization of 
WPI-induced destruction and quantum-erasure-induced revival of the interference
\textcolor{black}{
in the case of 
}%
unpolarized single photons
\textcolor{black}{
with their polarization states as the WWM.
}%

\begin{figure}[!t]
	\centering\includegraphics[scale=1, clip]{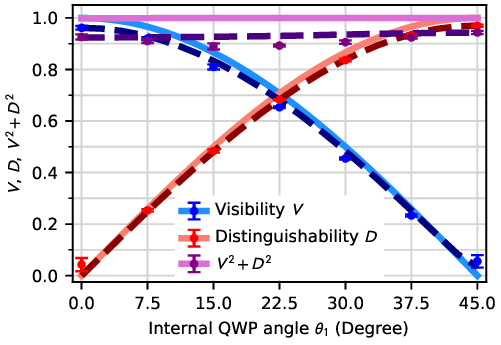}
	\caption{\label{fig:wpd} 
Comparison of the measured and theoretical values of the visibility $V$ (blue), the redefined path distinguishability $D$ (red), and $V^2+D^2$ (purple) as functions of $\theta_1$ when $\theta_0=0^\circ$. 
\textcolor{black}{The markers indicate the measured data, which are accompanied by the error bars that indicate 95\% confidence intervals. 
The 
solid curves indicate the ideal theoretical 
\textcolor{black}{
values.
}%
The 
dashed curves indicate the theoretical 
\textcolor{black}{
values taking account of
}%
the experimental 
\textcolor{black}{
imperfections 
(see text for details).
}%
}%
	}
\end{figure}

\subsection{\textcolor{black}{Verification of the WPD \textcolor{black}{equality}}}%
We 
\textcolor{black}{
examine
}%
the WPD 
\textcolor{black}{
equality, 
}%
\textcolor{black}{
Eq.~(\ref{eq:general.wpd}),
}%
by evaluating the visibility $V$ 
\textcolor{black}{
and
}%
the redefined path distinguishability $D$ as functions \textcolor{black}{of} $\theta_1$ from the complete absence of WPI ($\theta_1=0^\circ$) to the maximum presence of WPI ($\theta_1=45^\circ$) keeping $\theta_0=0^\circ$. 
\textcolor{black}{
The 
}%
\textcolor{black}{
visibility $V$ is measured by the same method as described in Sec.~4.2.
The
}%
distinguishability $D$ 
\textcolor{black}{
(Eqs.~(\ref{eq:general.mix.d_dcomp})--(\ref{eq:general.mix.d_dcomp2}))
}%
is measured as follows.
As described in 
Sec.~\ref{Sec.Theory},
\textcolor{black}{
let
$\bm{s}=p_{\alpha}\hat{\bm{s}}_{\alpha}+p_{\beta}\hat{\bm{s}}_{\beta}$,
}%
where
$\bm{s}$ is the Stokes vector of the input polarization state,
\textcolor{black}{
$\hat{\bm{s}}_{\alpha}$ and $\hat{\bm{s}}_{\beta}$
}%
are the Stokes vectors of the decomposed pure states 
that satisfy 
\textcolor{black}{
$\hat{\bm{n}}\cdot\hat{\bm{s}}_{\alpha}=\hat{\bm{n}}\cdot\hat{\bm{s}}_{\beta}=\hat{\bm{n}}\cdot\bm{s}$,
}%
\textcolor{black}{and} 
$\hat{\bm{n}}$ is the polarization rotation axis.
\textcolor{black}{
Then,
$D(\bm{s})$, i.e., $D$ for the input polarization \textcolor{black}{state} $\bm{s}$, 
is obtained as
\begin{align}
D(\bm{s})
&= p_{\alpha} D(\hat{\bm{s}}_\alpha) 
+ p_{\beta} D(\hat{\bm{s}}_\beta)
\label{eq:general.mix.d_dcomp2.1}
\\
&=
p_{\alpha}\max_{\bm{w}_\alpha} \left[2 L(\bm{w}_\alpha,\pi_{\alpha0},\pi_{\alpha1}) -1 \right]
+p_{\beta}\max_{\bm{w}_\beta} \left[2 L(\bm{w}_\beta,\pi_{\beta0},\pi_{\beta1}) -1 \right]
,
\label{eq:general.mix.d_dcomp2.2}
\end{align}
where
$D(\hat{\bm{s}}_\alpha)$ and $D(\hat{\bm{s}}_\beta)$,
i.e., 
$D$ for the pure states $\hat{\bm{s}}_\alpha$ and  $\hat{\bm{s}}_\beta$,
can be quantified using Eqs.~(\ref{eq:general.d_kmax}) and (\ref{eq:general.likelihood_c}).
}%
In our case where
$\hat{\bm{n}}=(0, 1, 0)$ 
and
$\bm{s}=\bm{0}$, 
the decomposed states can be 
any pair 
of orthogonal linear polarization states, e.g., 
$\ket{\rm H}$ 
($\hat{\bm{s}}_{\alpha}=(0, 0, 1)$)
and 
$\ket{\rm V}$  
($\hat{\bm{s}}_{\beta}=(0, 0, -1)$):
\begin{equation}
D(\bm{s}=\bm{0})=
\textcolor{black}{
\frac12 D (\hat{\bm{s}}_{\alpha}=(0, 0, 1))+
\frac12 D (\hat{\bm{s}}_{\beta}=(0, 0, -1)). 
}%
\label{eq:dhv}
\end{equation}
This means 
that
we can estimate $D(\bm{s}=\bm{0})$ by measuring the 
distinguishability 
\textcolor{black}{
for the decomposed pure states
}%
\textcolor{black}{
$D(\hat{\bm{s}}_{\alpha})$ and $D(\hat{\bm{s}}_{\beta})$ 
}%
with 
a
polarizer 
after the photon source 
in order
to pass 
either
$\ket{\rm H}$ or $\ket{\rm V}$ 
component of the 
unpolarized photons. 
In practice,
the Stokes vector
$\bm{s}$ of our photon source deviates
slightly from $\bm{0}$
and is approximated as $\bm{s}\simeq(0, 0, s_3)$.
Thus, we used the decomposition
\begin{equation}
D(\bm{s}=(0, 0, s_3))=
\textcolor{black}{
p_{\alpha}D(\hat{\bm{s}}_{\alpha}=(0, 0, 1))+
p_{\beta}D(\hat{\bm{s}}_{\beta}=(0, 0, -1)),  
}%
\label{eq:dhv2}
\end{equation}
where 
$\textcolor{black}{p_{\alpha}}=\frac12(1+s_3)$ and $\textcolor{black}{p_{\beta}}=\frac12(1-s_3)$.  
Thus, we can obtain 
$D(\bm{s})$ 
by evaluating 
\textcolor{black}{%
the right hand side of
}%
Eq.~(\ref{eq:dhv2}). 
The 
distinguishability 
\textcolor{black}{
for the pure states
}%
is evaluated
using the likelihood of 
which-way
\textcolor{black}{
guess,
$L (\bm{w},\pi_0,\pi_1)$
given in
Eq.~(\ref{eq:general.likelihood_c}) 
}%
\cite{schwindt1999, englert1999, englert1996}. 
\textcolor{black}{
In the experiment,
the polarization state $\pi_0$ ($\pi_1$) 
of a photon passing through path 0 (1)
is selected by keeping path 0 (1) open
while the other path is blocked.
The detection basis set 
$\bm{w}$
of the output polarization state
is chosen by adjusting HWP1 ($\theta_\mathrm{H1}$), QWP3 ($\theta_\mathrm{3}$) and PBS
in the output path 1,
and the outcome is detected on either APD\textcolor{black}{10} or APD\textcolor{black}{11}. 
}%
\textcolor{black}{
Thus,
}%
the 
likelihood  
\textcolor{black}{
$L$
for the input pure polarization state ($\hat{\bm{s}}_{\alpha}$ or $\hat{\bm{s}}_{\beta}$) 
}%
is  
quantified by
\begin{equation}
\textcolor{black}{
L
}%
=\frac{\max \{N_{0, 10}, N_{1, 10}\}+\max \{N_{0, 11}, N_{1, 11}\}}{N_{0, 10}+N_{1, 10}+N_{0, 11}+N_{1, 11}}, 
\label{eq:l}
\end{equation} 
where $N_{p, d}$  is the 
count rate of APD$d$ ($d=10, 11$) 
when path $p$ ($p=0, 1$) is open whereas the other path is blocked. 
It is known 
that 
the optimal polarization detection bases 
that maximize  
\textcolor{black}{
$L$
}%
are achieved by
setting
$\theta_3=0^\circ$
and
$\theta_{\rm H1}=-\theta_1/2+22.5^\circ$ 
($\theta_{\rm H1}=-\theta_1/2+67.5^\circ$)
for 
the input polarization of 
$\hat{\bm{s}}_{\alpha}=(0, 0, 1)$
($\hat{\bm{s}}_{\beta}=(0, 0, -1)$)  
\cite{schwindt1999}%
.
\textcolor{black}{%
Using $L$ thus obtained,
we experimentally quantify 
$D(\hat{\bm{s}}_{\alpha})$, $D(\hat{\bm{s}}_{\beta})$, and $D(\bm{s})$.
}%
The measured data of $V$, $D$, and $V^2+D^2$ are shown in Fig.~\ref{fig:wpd} along with their \textcolor{black}{ideal theoretical curves (Eqs.~(\ref{eq:vnews}), (\ref{eq:dnews}), and (\ref{eq:wpdnews}), respectively) and their theoretical curves 
\textcolor{black}{
taking
}%
experimental imperfections
\textcolor{black}{
into account.
}%
Here, 
\textcolor{black}{
the
}%
imperfections 
\textcolor{black}{
include
}%
the input polarization state
\textcolor{black}{
that deviates slightly from the completely unpolarized state,
}%
the 
\textcolor{black}{
dispersive
error in the
}%
retardance of the wave plates, 
\textcolor{black}{the }%
\textcolor{black}{%
polarization-dependent 
imbalance in the
transmittance/reflectance
}%
of the 
\textcolor{black}{NPBS, 
the 
imperfect extinction ratios 
of the PBSs, }%
and 
\textcolor{black}{%
the 
polarization-dependent
}%
reflectance 
of the retroreflectors.
The amounts and effects of 
these errors and imperfections are estimated, when possible, 
as weighted averages over the filtered emission spectrum of the photon source.
In addition, 
\textcolor{black}{
degradation of 
}%
overall visibility 
due to 
\textcolor{black}{
possible
}%
wavefront 
\textcolor{black}{
distortion
}%
and misalignment
\textcolor{black}{
is considered.
}%
The measured data are in reasonable agreement with the theoretical curves
\textcolor{black}{
taking account of
}%
the experimental imperfections. 
We note that the error bars deal only with statistical errors 
\textcolor{black}{
originating from detected photon number fluctuations
}%
and do not 
\textcolor{black}{
include
}%
systematic errors 
due to 
\textcolor{black}{
the
}%
experimental imperfections. 
The remaining deviation
\textcolor{black}{
between the experiment and the theory
}%
might be attributed to 
\textcolor{black}{
additional
}%
experimental imperfections
that are not yet taken into account.
}%
Despite the slight difference, the values of $V^2+D^2$ obtained from the measurement are close to unity, demonstrating that $V$ and $D$ thus measured almost satisfy the WPD equality.
Thus, we have experimentally verified the complementary relation between the visibility $V$ and the redefined distinguishability $D$; 
they in principle satisfy the WPD equality for all pure and mixed polarization states. 

\section{Concluding remarks}
In this paper, we have investigated 
the WPD relation associated with polarization-based WPI in unpolarized photons.
\textcolor{black}{
We have introduced the generalized measure of 
}%
path distinguishability $D$, 
\textcolor{black}{
so that
}%
the WPD equality with the visibility $V$, i.e., $V^2+D^2=1$ (Eq.~(\ref{eq:general.wpd})), holds for any polarization state including the unpolarized state within a purification-based framework,
whereas the conventional distinguishability $D_\mathrm{c}$ exhibits the WPD inequality $V^2+D_\mathrm{c}^2\le1$, where the equality holds only for pure polarization states. 
\textcolor{black}{
We have shown that 
the WPI in the case of mixed-state WWM 
is not only held in the WWM alone 
but also shared between the WWM and an external system, 
e.g., environment, entangled with the WWM.
Thus, 
$D$ must
incorporate the WPI not only held in the WWM alone 
but also shared through the entanglement.
To this end, 
we proposed the appropriate and practical measure of $D$,
in terms of optimal decomposition of the mixed state into a couple of pure states
that are intrinsically distinguishable with each other because of the entanglement.
$D$ thus defined is experimentally quantifiable 
even if the system entangled with the WWM is 
no longer 
accessible.
With $D$, 
we can now test the perfect complementary relation, i.e\textcolor{black}{.}, WPD equality, 
even if
the WWM is in a mixed state. 
}%

\textcolor{black}{
In the experiment,
using true single photons 
in the completely mixed (unpolarized) state emitted from an NV center in diamond,
}%
we have demonstrated the WPI-induced~destruction and the
quantum-erasure-induced~revival of the interference of 
unpolarized 
photons.
The interference vanishes at the maximum WPI even though the polarization state, i.e., the completely unpolarized state, is unchanged and identical in each arm.
Quantum-erasure-induced revival of the interference appears in the same manner as in the case of photons in pure polarization states. 
Then, most importantly, we have experimentally investigated the WPD relation and verified $V^2+D^2\simeq1$ within slight experimental errors.
\textcolor{black}{This} is \textcolor{black}{an} 
experimental demonstration that any \textcolor{black}{polarization-based} WWM 
not only in a pure state but also in a mixed state 
can exhibit the perfect complementary relation in terms of $V$ and $D$.   

Wave--particle duality and complementarity are fundamentals in quantum physics. 
Our experimental results have demonstrated that full path-distinguishability can be added even if the WWM is in a mixed qubit state, and the distinguishability, i.e., the particle nature, is properly quantified by $D$, which presents the full complementarity with the visibility $V$, i.e., the wave nature. 
We note that what we proposed and demonstrated here are limited to simple systems where a qubit, such like photon polarization, is used as a WWM.
Although our definition of the distinguishability $D$ 
\textcolor{black}{
(Eqs.~(\ref{eq:general.mix.d_kmax})--(\ref{eq:general.mix.d_dcomp2}))
}%
can in principle be generalized to higher dimensional systems, to find a set of 
\textcolor{black}{
optimal
}%
decomposition that 
\textcolor{black}{leads to $D$}
would be 
more difficult as increasing degrees of freedom. 
In this regard, PCT in multi\textcolor{black}{-}path scenarios \cite{Qureshi2021Predictability, Paul2020Multipath}
\textcolor{black}{is} 
worth noting. 
Nevertheless, our findings will be useful in more complete understanding of the WPI and the distinguishability added by the WWM encoded in general qubit systems including mixed states.   
We also hope that our work will stimulate subsequent studies and discussions that aim to gain further insights into the WPD and complementarity, in the light of fundamental quantum physics and quantum optics.

\section*{Funding. }
\textcolor{black}{
MEXT Q-LEAP (No.~JPMXS0118067581).
}%

\section*{Acknowledgments. }
The authors greatly appreciate fruitful discussions with 
Mark Sadgrove, Yasuyoshi Mitsumori\textcolor{black}{, 
Fumihiro Kaneda, and Soyoung Baek}. 
N. A. thanks Tohoku University Division for Interdisciplinary Advanced Research and Education for financial support.
This research was partly conducted at the Fundamental Technology Center, Research Institute of Electrical Communication, Tohoku University. 

\section*{Disclosures. }
The authors declare no conflicts of interest.

\section*{Data availability. }
Data underlying the results presented in this paper are not publicly available at this time but may be obtained from the authors upon reasonable request. 

\section*{Supplemental document. }
See Supplement 1 for supporting content.

%
%
%

%
%
%
%
%
%
%
%

%
%
%
%
%
%
%

%
%


\begin{thebibliography}{10}
\newcommand{\enquote}[1]{``#1''}

\bibitem{jacques2005}
V.~Jacques, E.~Wu, T.~Toury, \emph{et~al.}, \enquote{Single-photon
  wavefront-splitting interference,} {\protect\JournalTitle{Eur. Phys. J. D}}
  \textbf{35}, 561--565 (2005).

\bibitem{xiao2019observing}
Y.~Xiao, H.~M. Wiseman, J.-S. Xu, \emph{et~al.}, \enquote{Observing momentum
  disturbance in double-slit `which-way' measurements,}
  {\protect\JournalTitle{Science Advances}} \textbf{5}, eaav9547 (2019).

\bibitem{bach2013}
R.~Bach, D.~Pope, S.-H. Liou, and H.~Batelaan, \enquote{Controlled double-slit
  electron diffraction,} {\protect\JournalTitle{New Journal of Physics}}
  \textbf{15}, 033018 (2013).

\bibitem{zeilinger1988}
A.~Zeilinger, R.~G{\"a}hler, C.~Shull, \emph{et~al.}, \enquote{Single-and
  double-slit diffraction of neutrons,} {\protect\JournalTitle{Reviews of
  modern physics}} \textbf{60}, 1067--1073 (1988).

\bibitem{carnal1991}
O.~Carnal and J.~Mlynek, \enquote{Young's double-slit experiment with atoms: A
  simple atom interferometer,} {\protect\JournalTitle{Physical review letters}}
  \textbf{66}, 2689--2692 (1991).

\bibitem{nairz2003}
O.~Nairz, M.~Arndt, and A.~Zeilinger, \enquote{Quantum interference experiments
  with large molecules,} {\protect\JournalTitle{American Journal of Physics}}
  \textbf{71}, 319--325 (2003).

\bibitem{englert1996}
B.-G. Englert, \enquote{Fringe visibility and which-way information: An
  inequality,} {\protect\JournalTitle{Physical review letters}} \textbf{77},
  2154--2157 (1996).

\bibitem{englert1999}
B.-G. Englert, \enquote{Remarks on some basic issues in quantum mechanics,}
  {\protect\JournalTitle{Zeitschrift f{\"u}r Naturforschung A}} \textbf{54},
  11--32 (1999).

\bibitem{bosyk2013}
G.~Bosyk, M.~Portesi, F.~Holik, and A.~Plastino, \enquote{On the connection
  between complementarity and uncertainty principles in the {M}ach-{Z}ehnder
  interferometric setting,} {\protect\JournalTitle{Physica Scripta}}
  \textbf{87}, 065002 (2013).

\bibitem{zela2018}
F.~De~Zela, \enquote{Hidden coherences and two-state systems,}
  {\protect\JournalTitle{Optica}} \textbf{5}, 243--250 (2018).

\bibitem{jara2022experimental}
M.~Jara, J.~Marrou, M.~Uria, \emph{et~al.}, \enquote{Experimental display of
  generalized wave-particle duality,} {\protect\JournalTitle{Optics Express}}
  \textbf{30}, 34740--34749 (2022).

\bibitem{sanchez2019}
P.~S{\'a}nchez, J.~Gonzales, V.~Avalos, \emph{et~al.}, \enquote{Experimental
  display of the extended polarization coherence theorem,}
  {\protect\JournalTitle{Optics letters}} \textbf{44}, 1052--1055 (2019).

\bibitem{schwindt1999}
P.~D. Schwindt, P.~G. Kwiat, and B.-G. Englert, \enquote{Quantitative
  wave-particle duality and nonerasing quantum erasure,}
  {\protect\JournalTitle{Physical Review A}} \textbf{60}, 4285--4290 (1999).

\bibitem{liu2012}
H.-Y. Liu, J.-H. Huang, J.-R. Gao, \emph{et~al.}, \enquote{Relation between
  wave-particle duality and quantum uncertainty,}
  {\protect\JournalTitle{Physical Review A}} \textbf{85}, 022106 (2012).

\bibitem{jacques2007}
V.~Jacques, E.~Wu, F.~Grosshans, \emph{et~al.}, \enquote{Experimental
  realization of {W}heeler{\textquoteright}s delayed-choice gedanken
  experiment,} {\protect\JournalTitle{Science}} \textbf{315}, 966--968 (2007).

\bibitem{jacques2008delayed}
V.~Jacques, E.~Wu, F.~Grosshans, \emph{et~al.}, \enquote{Delayed-choice test of
  quantum complementarity with interfering single photons,}
  {\protect\JournalTitle{Physical Review Letters}} \textbf{100}, 220402 (2008).

\bibitem{kaiser2012}
F.~Kaiser, T.~Coudreau, P.~Milman, \emph{et~al.}, \enquote{Entanglement-enabled
  delayed-choice experiment,} {\protect\JournalTitle{Science}} \textbf{338},
  637--640 (2012).

\bibitem{ma2016}
X.-s. Ma, J.~Kofler, and A.~Zeilinger, \enquote{Delayed-choice gedanken
  experiments and their realizations,} {\protect\JournalTitle{Reviews of Modern
  Physics}} \textbf{88}, 015005 (2016).

\bibitem{steuernagel2007}
O.~Steuernagel, \enquote{Afshar's experiment does not show a violation of
  complementarity,} {\protect\JournalTitle{Foundations of Physics}}
  \textbf{37}, 1370--1385 (2007).

\bibitem{jacques2008illustration}
V.~Jacques, N.~Lai, A.~Dr{\'e}au, \emph{et~al.}, \enquote{Illustration of
  quantum complementarity using single photons interfering on a grating,}
  {\protect\JournalTitle{New Journal of Physics}} \textbf{10}, 123009 (2008).

\bibitem{durr1998}
S.~D{\"u}rr, T.~Nonn, and G.~Rempe, \enquote{Fringe visibility and which-way
  information in an atom interferometer,} {\protect\JournalTitle{Physical
  review letters}} \textbf{81}, 5705--5709 (1998).

\bibitem{ericsson2005}
M.~Ericsson, D.~Achilles, J.~T. Barreiro, \emph{et~al.}, \enquote{Measurement
  of geometric phase for mixed states using single photon interferometry,}
  {\protect\JournalTitle{Phys. Rev. Lett.}} \textbf{94}, 050401 (2005).

\bibitem{Qureshi2021Predictability}
T.~Qureshi, \enquote{Predictability, distinguishability, and entanglement,}
  {\protect\JournalTitle{Opt. Lett.}} \textbf{46}, 492--495 (2021).

\bibitem{abe2017}
N.~Abe, Y.~Mitsumori, M.~Sadgrove, and K.~Edamatsu, \enquote{Dynamically
  unpolarized single-photon source in diamond with intrinsic randomness,}
  {\protect\JournalTitle{Scientific reports}} \textbf{7}, 46722 (2017).

\bibitem{eberly2017}
J.~Eberly, X.-F. Qian, and A.~Vamivakas, \enquote{Polarization coherence
  theorem,} {\protect\JournalTitle{Optica}} \textbf{4}, 1113--1114 (2017).

\bibitem{qian2018}
X.-F. Qian, A.~Vamivakas, and J.~Eberly, \enquote{Entanglement limits duality
  and vice versa,} {\protect\JournalTitle{Optica}} \textbf{5}, 942--947 (2018).

\bibitem{Kanseri2018Experimental}
B.~Kanseri and K.~R. Sethuraj, \enquote{Experimental observation of the
  polarization coherence theorem,} {\protect\JournalTitle{Opt. Lett.}}
  \textbf{44}, 159--162 (2019).

\bibitem{Qian2020Turning}
X.-F. Qian, K.~Konthasinghe, S.~K. Manikandan, \emph{et~al.}, \enquote{Turning
  off quantum duality,} {\protect\JournalTitle{Phys. Rev. Res.}} \textbf{2},
  012016 (2020).

\bibitem{Hill1997Entanglement}
S.~A. Hill and W.~K. Wootters, \enquote{Entanglement of a pair of quantum
  bits,} {\protect\JournalTitle{Phys. Rev. Lett.}} \textbf{78}, 5022--5025
  (1997).

\bibitem{Wootters1998Entanglement}
W.~K. Wootters, \enquote{Entanglement of formation of an arbitrary state of two
  qubits,} {\protect\JournalTitle{Phys. Rev. Lett.}} \textbf{80}, 2245--2248
  (1998).

\bibitem{Rungta2001Universal}
P.~Rungta, V.~Bu\ifmmode~\check{z}\else \v{z}\fi{}ek, C.~M. Caves,
  \emph{et~al.}, \enquote{Universal state inversion and concurrence in
  arbitrary dimensions,} {\protect\JournalTitle{Phys. Rev. A}} \textbf{64},
  042315 (2001).

\bibitem{Torre2022Relationship}
C.~R.~M. Montenegro La~Torre, Y.~Yugra, and F.~De~Zela, \enquote{Relationship
  between entanglement and polarization in tripartite states,}
  {\protect\JournalTitle{Journal of Optics}} \textbf{24}, 105202 (2022).

\bibitem{Marrou2023Wave}
J.~P. Marrou, C.~M.~L. Torre, M.~Jara, and F.~D. Zela, \enquote{Wave--particle
  duality in tripartite systems,} {\protect\JournalTitle{J. Opt. Soc. Am. A}}
  \textbf{40}, C22--C29 (2023).

\bibitem{Qian2020Quantification}
X.-F. Qian, S.~A. Wadood, A.~N. Vamivakas, and J.~H. Eberly,
  \enquote{Quantification and observation of genuine three-party coherence: A
  solution based on classical optics,} {\protect\JournalTitle{Phys. Rev. A}}
  \textbf{102}, 062424 (2020).

\bibitem{Qian2016Coherence}
X.-F. Qian, T.~Malhotra, A.~N. Vamivakas, and J.~H. Eberly, \enquote{Coherence
  constraints and the last hidden optical coherence,}
  {\protect\JournalTitle{Phys. Rev. Lett.}} \textbf{117}, 153901 (2016).

\bibitem{peters2003}
N.~Peters, J.~Altepeter, E.~Jeffrey, \emph{et~al.}, \enquote{Precise creation,
  characterization, and manipulation of single optical qubits,}
  {\protect\JournalTitle{Quantum Inf. Comput.}} \textbf{3}, 503--517 (2003).

\bibitem{hertel2015}
I.~V. Hertel and C.-P. Schulz, \emph{Atoms, Molecules and Optical Physics 2}
  (Springer, 2015), Chap.~2.

\bibitem{jelezko2003}
F.~Jelezko, A.~Volkmer, I.~Popa, \emph{et~al.}, \enquote{Coherence length of
  photons from a single quantum system,} {\protect\JournalTitle{Physical Review
  A}} \textbf{67}, 041802 (2003).

\bibitem{braig2003}
C.~Braig, P.~Zarda, C.~Kurtsiefer, and H.~Weinfurter, \enquote{Experimental
  demonstration of complementarity with single photons,}
  {\protect\JournalTitle{Applied Physics B}} \textbf{76}, 113--116 (2003).

\bibitem{marshall2011}
G.~D. Marshall, T.~Gaebel, J.~C. Matthews, \emph{et~al.}, \enquote{Coherence
  properties of a single dipole emitter in diamond,} {\protect\JournalTitle{New
  Journal of Physics}} \textbf{13}, 055016 (2011).

\bibitem{kim2000}
Y.-H. Kim, R.~Yu, S.~P. Kulik, \emph{et~al.}, \enquote{Delayed ``choice''
  quantum eraser,} {\protect\JournalTitle{Physical Review Letters}}
  \textbf{84}, 1--5 (2000).

\bibitem{walborn2002}
S.~Walborn, M.~T. Cunha, S.~P{\'a}dua, and C.~Monken, \enquote{Double-slit
  quantum eraser,} {\protect\JournalTitle{Physical Review A}} \textbf{65},
  033818 (2002).

\bibitem{gogo2005}
A.~Gogo, W.~D. Snyder, and M.~Beck, \enquote{Comparing quantum and classical
  correlations in a quantum eraser,} {\protect\JournalTitle{Physical Review A}}
  \textbf{71}, 052103 (2005).

\bibitem{Paul2020Multipath}
B.~Paul, S.~Kamal, and T.~Qureshi, \enquote{Multipath wave-particle duality in
  classical optics,} {\protect\JournalTitle{Opt. Lett.}} \textbf{45},
  3204--3207 (2020).

\end{thebibliography}
\providecommand{\noopsort}[1]{}\providecommand{\singleletter}[1]{#1}%

\end{document}